\documentclass[aps,prx,nofootinbib,twocolumn,amsmath,amssymb,superscriptaddress,notitlepage,llncs]{revtex4-1}

\usepackage{hyperref}
\usepackage{soul}
\usepackage{latexsym}
\usepackage{graphicx}
\usepackage{times,psfrag,subfigure}
\usepackage{enumerate}
\usepackage{amsmath}
\usepackage{dsfont}
\usepackage{dcolumn}
\usepackage{bm,bbm}
\usepackage{color}
\usepackage{latexsym,amsmath,amssymb,bm,euscript}
\usepackage{dsfont}
\usepackage{textcomp}
\usepackage{tabularx}
\usepackage{setspace}
\usepackage{ctable}
\usepackage{sidecap}
\usepackage{placeins}
\usepackage{threeparttable}
\usepackage{ gensymb,wasysym}
\usepackage{epsfig,subfigure}
\usepackage{amsfonts}
\usepackage{physics}
\usepackage{exscale}
\usepackage{amsbsy}

\let \oldbm \bm
\renewcommand{\vec}[1]{\oldbm{#1}}

\def\bk{{\vec k}}
\def\bp{{\vec p}}
\def\bg{{\vec g}}
\def\be{{\vec e}}
\def\bA{{\vec A}}
\def\bB{{\vec B}}
\def\bz{{\vec z}}
\def\bS{{\vec S}}

\def\bz{{\vec z}}
\def\bK{{\vec K}}
\def\bq{{\vec q}}

\def\bR{{\vec R}}
\def\bG{{\vec G}}
\def\bd{{\vec d}}

\def\bm{{\vec m}}
\def\br{{\vec r}}

\def\bdelta{{\boldsymbol \delta}}

\def\bsigma{{\boldsymbol \sigma}}

\def\bs{\bsigma}

\def\c1{c$_1$}
\def\c2{c$_2$}
\def\v1{v$_1$}
\def\v2{v$_2$}

\def\tr{\mathop{\mathrm{tr}}}

\def\T{\mathcal{T}}

\def\H{\mathcal{H}}
\def\K{\mathcal{K}}

\def\diag{{\rm diag}}

\newcommand{\beq}{\begin{equation}}
\newcommand{\eeq}{\end{equation}}
\newcommand{\beqarray}{\begin{eqnarray}}
\newcommand{\eeqarray}{\end{eqnarray}}
\newcommand{\change}[1]{{\color{black} #1}}

\newcommand{\printfnsymbol}[1]{%
  \textsuperscript{\@fnsymbol{#1}}%
}

\allowdisplaybreaks

\graphicspath{{figures/}}

\begin{document}

\title{
Theory of correlated insulating behaviour and spin-triplet superconductivity in twisted double bilayer graphene}

\author{Jong Yeon Lee}
\thanks{J.Y. Lee and E. Khalaf contributed equally to this work.}
\author{Eslam Khalaf}
\thanks{J.Y. Lee and E. Khalaf contributed equally to this work.}
\author{Shang Liu}
\author{Xiaomeng Liu}
\author{Zeyu Hao}
\author{Philip Kim}
\author{Ashvin Vishwanath}
\affiliation{Department of Physics, Harvard University, Cambridge, Massachusetts 02138, USA}

\begin{abstract}
Two monolayers of graphene twisted by a small `magic' angle  exhibit nearly flat bands leading to correlated electronic states and superconductivity, whose precise nature, e.g. possible broken symmetries, remain under debate.  Here we theoretically  study  a related but different system with reduced symmetry - twisted {\em double} bilayer graphene (TDBG), consisting of {\em two} Bernal stacked bilayer graphene sheets, twisted with respect to one another. Unlike the monolayer case, we show that isolated  flat bands only appear on application of a vertical displacement field $D$.  We construct a phase diagram as a function of twist angle and $D$, incorporating interactions via a Hartree-Fock approximation. At half filling, ferromagnetic insulators are stabilized, typically with valley Chern number $C_v= \pm 2$. Ferromagnetic fluctuations in the metallic state are argued to lead to spin triplet superconductivity from pairing between electrons in opposite valleys. Response of these states to a magnetic field applied either perpendicular or parallel to the graphene sheets is obtained, and found to compare favorably with a recent experiment. We highlight a novel orbital effect arising from in-plane fields that is comparable to the Zeeman effect and plays an important role in interpreting experiments. Combined with recent experimental findings, our results establish TDBG as a tunable platform to realize rare phases in conventional solids, such as ferromagnetic insulators and spin-triplet superconductors.
\end{abstract}

\maketitle

\section{Introduction}

The recent discovery of correlated insulating states and superconductivity in twisted bilayer graphene (TBG) \cite{PabloMott, PabloSC, Dean-Young, Efetov} has opened a new window to exploring strong correlation effects in systems whose doping can be easily tuned, enabling the exploration of a rich range of interaction-driven phenomena. Although the underlying reason for the correlated physics is understood to arise from a relatively narrow electronic bandwidth induced by the long wavelength Moir\'e pattern \cite{MacDonald, Santos}, several details, including the symmetry breaking within the insulating phase and the nature and mechanism of pairing in the neighboring superconductor, remain under debate \cite{Balents18, Po2018, IsobeFu, Thomson18, YouAV, Vafek, Xie2018,Nandkishore, Kivelson, PhilipPhillips,phononMacDonald, phononLianBernevig, Zou2018}. One of the difficulties in addressing these questions arises from the complexity of the theoretical treatment of TBG which involves at least a pair of narrow bands per spin per  valley with a symmetry-protected band touching, leading to 8 bands in total. On top of that, the limited tunability of the band structure makes it experimentally difficult to explore the dependence of different phases on microscopic parameters.

Motivated by recent experimental report \cite{TDBGexp2019}, we study a related system--- twisted double bilayer graphene (TDBG)--- which consists of a {\em pair} of bilayer graphene sheets, twisted with respect to one another with AB-AB stacking structure. Due to the absence of $C_2$ rotation symmetry, TDBG has a lower symmetry compared to TBG which simplifies the problem by removing the band touching at the Dirac points, leading to a low energy effective description involving one rather than two narrow bands per spin and valley. Moreover, the band separation can be controlled by applying a vertical displacement field enabling the exploration of different regimes of band isolation and bandwidth within the same device.

\begin{figure}[t]
\begin{center}
\includegraphics[width=0.7\columnwidth]{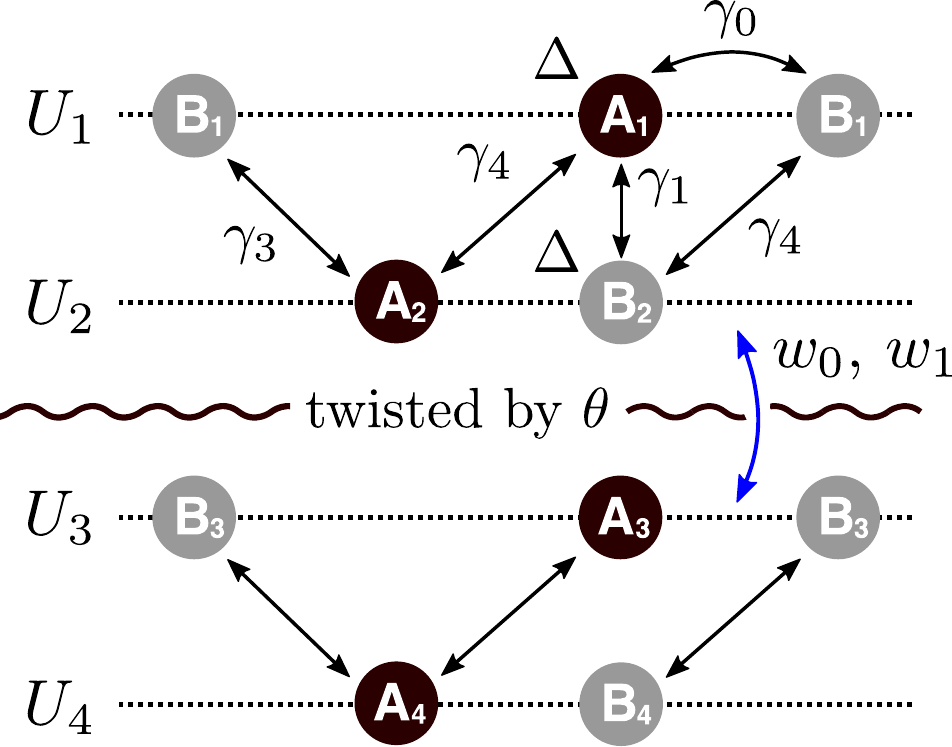}
\caption{Twisted double BLG model (ABAB stacking) with the gating voltage $U$ across the system. Throughout this work, we assume the voltage drop across the layers is uniform, $U_i-U_{i+1} = U/3$. }
\label{fig:ABAB_schematic}
\end{center}
\end{figure}

We identify three main ingredients necessary to explain the emergence of insulating and superconducting behavior in TDBG. First, we perform an accurate calculation of the single-particle band structure to identify ranges of displacement field and twist angle for which a single band is isolated and relatively flat. We show that lattice relaxation, known to be important in TBG \cite{Koshino2017, Koshino2018}, as well as several other effects such as trigonal warping, which are absent in TBG, significantly influence the band structure in TDBG, in excellent agreement with experiments.
Moreover, we identify a hitherto neglected in-plane orbital effect which is used to explain the experimentally observed deviation of the in-plane $g$ factor from 2 \cite{TDBGexp2019}, as well as the effect of in-plane field on superconducting $T_c$.

Second, we address the key question of the nature of the interaction-driven insulating state. The similarity between the phase diagram of TBG to that of cuprates was invoked to argue that Mott physics is the underlying mechanism responsible for the correlated insulator \cite{PabloMott, Balents18, Vafek}. On the other hand, a different route to correlated insulators is observed in graphene quantum Hall systems, for instance, when the spin and valley degeneracy of the Landau levels are spontaneously broken by interactions \cite{Goerbig}. This usually leads to ferromagnetic insulators, which are otherwise rare in correlated solids where antiferromagnetic order is the norm. 
For similar reasons, in the TDBG with  non-zero valley Chern number, ferromagnetism may be preferred \cite{Zhang2018} at integer fillings. The situation here is reminiscent of strained graphene, where a suitably chosen strain profile leads to Landau levels arising from the opposite strain magnetic fields applied on the two valleys \cite{Crommie2010}. At partial fillings that are integers, ferromagnetic ground states were obtained with repulsive interactions \cite{Ghaemi2012}, and we show that a similar scenario is likely to occur here in TDBG. Indeed a related ground state with spontaneous quantum Hall response, although metallic, was observed in the twisted monolayer-monolayer graphene (TBG) with $C_2$-breaking substrate potentials \cite{Sharpe2019, Xie2018, Zou2018, Zhang2018, Zhang2019, Bultinck19}.

Third, we investigate the nature of the superconducting phase by highlighting that the valley degree of freedom, which behaves as a pseudo-spin, allows for exotic pairing possibilities which are relatively rare in other materials. In particular, we show that spin-triplet with valley-singlet pairing, which is momentum-independent within each valley, is favored. We investigate the consequences of such scenario and show it can be used to explain the measured dependence of $T_c$ on in-plane field \cite{TDBGexp2019}.

\begin{figure}[t]
\begin{center}
\includegraphics[width=0.9\columnwidth]{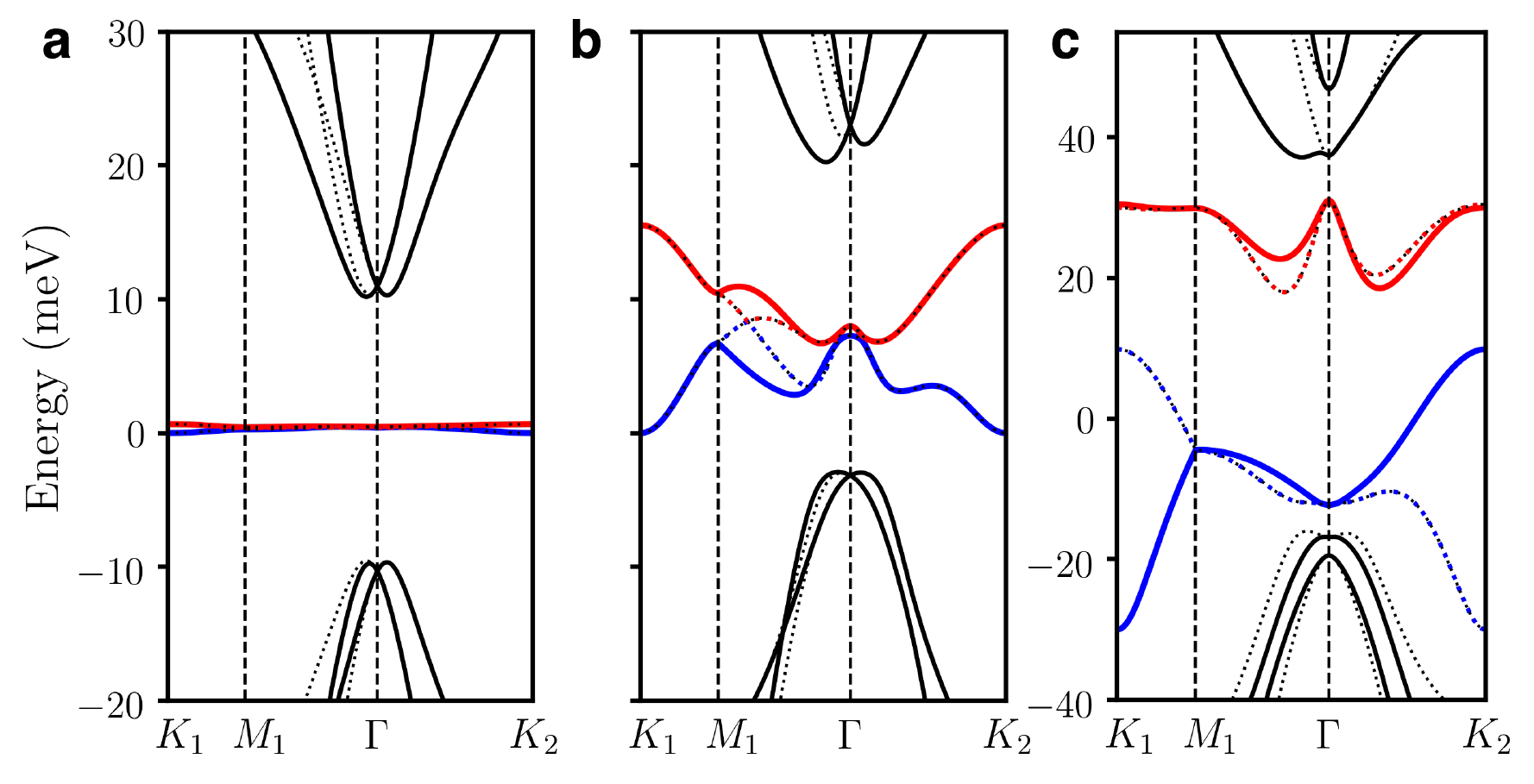}
\caption{
\textbf{a,b} Moir\'e band structures at $(\theta,U) = (1.05^\circ,0)$. Solid (dotted) line represesnts the band originated from $\bK_+$ ($\bK_-$) valley. Red, blue and black represent conduction, valence, and the other bands respectively. \textbf{a,} The band structure for the idealized model with only $\gamma_0$ and $\gamma_1$ being non-zero. The flat band is observed with the bandwidth $0.25$\,meV.  \textbf{b,} The band structure for the realistic model with overlapping bands. The `magic angle' does not exist in this case. \textbf{c,} Moir\'e band structure at $(\theta,U) = (1.33^{\circ},60\textrm{ meV})$. The first conduction band (red) is isolated and relatively flat.  }
\label{fig:band_trigonal}
\end{center}
\end{figure}

\section{Results}
\subsection{Single-particle physics}
\label{sec:single_particle}

We consider a system consisting of two AB-stacked graphene bilayers twisted relative to ABAB stacking by a small angle $\theta$, illustrated in Fig.\,\ref{fig:ABAB_schematic}. For a detailed discussion on the Hamiltonian and model parameters, see Methods. The bottom layer of the top BLG and the top layer of the bottom BLG are coupled via Moir\'e hopping between $AA$ and $AB$ sites, parametrized by $(w_0, w_1)$ \cite{Koshino2017, Koshino2018}. In the original Bistritzer-Macdonald model, $w_0$ and $w_1$ are taken to be equal \cite{MacDonald2011}. However, in a realistic twisted model, the ratio $r \equiv w_0/w_1$ is smaller than one due to the lattice relaxation which expands (shrinks) AB (AA) regions. In TBG, $r$ is taken to be around 0.75 for the first magic angle \cite{Koshino2017, Koshino2018}. Here, we similarly include lattice relaxations by taking $r$ to be smaller than 1. This is crucial for the existence of a gap between first and second conduction (valence) bands in TDBG which is necessary to explain the band insulator at $\nu= \pm 4$ filling. In this work, we take $(w_0,w_1) = (88,100)\textrm{ meV}$ corresponding to $r=0.88$. For different values of $(w_0,w_1)$, we obtained qualitatively similar features (Methods).

\begin{figure*}[t]
\begin{center}
\includegraphics[width=0.92\textwidth]{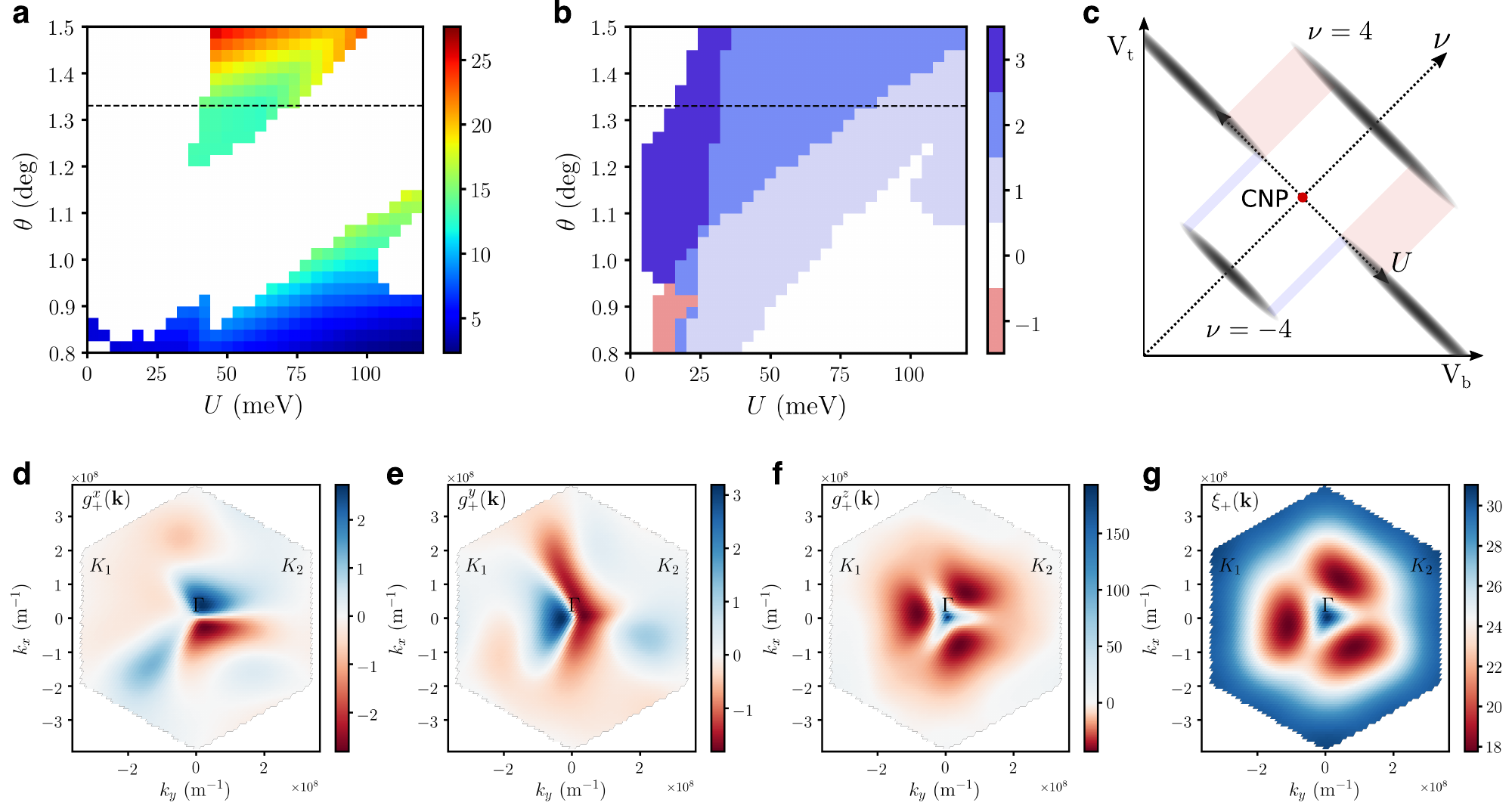}

\vspace{-5pt}

\caption{ \textbf{a,} Isolation region for the first conduction band (colored) with the bandwith indicated by the color. We observe two seperate isolation regions for $\theta$ smaller or larger than $1.1^\circ$. The former is not very robust and is sensitive to fine-tuning of parameters whereas the latter is very robust and is associated with a valley Chern number of 2 (See \textbf{b}).  
\textbf{b,} The Chern number of the first conduction band from $\bK_+$ valley. Note, the Chern number is defined as long as a {\em direct} band gap is present. \textbf{c,} A schematic plot for the insulating (black) regions and the first conduction/valence band isolated region (red/blue) in the TDBG at $\theta=1.33^\circ$. The red dot is charge neutrality point (CNP). In the shaded region, strongly correlated physics is expected near integer-fillings. Asymmetry between electron and hole dopings is predicted from the theory.
\textbf{d, e, f, g}, Color plots for $g$-factor associated with orbital magnetic effects $g^x_+(\bk)$,  $g^y_+(\bk)$,  $g^z_+(\bk)$, and single-particle dispersion $\xi_+(\bk)$ over the Moir\'e Brillouin zone for the first conduction band at $(\theta,U) = (1.33^\circ, 60 \textrm{ meV})$, where the band is isolated. $g^{x,y,z}(\bk)$ are in the unit of $\mu_B$, and $\xi(\bk)$ is in the unit of meV. Both $g^x$ and $g^y$ vanish at high symmetric points $\Gamma$, $K_1$ and $K_2$. 
}
\label{fig:single_particle}
\end{center}
\end{figure*}

Unlike TBG, a realistic description of TDBG does not exhibit magic angle physics whose origin is the vicinity to a chiral symmetric model with perfectly flat bands at specific angles \cite{Tarnopolsky, Khalaf2019}. In the quadratic approximation of the bilayer-graphene dispersion, the first conduction and valence bands in TDBG become almost perfectly flat at the angle $\theta \approx 1.05$ \cite{Zhang2018}. However, once trigonal warping ($\gamma_3$) and particle-hole asymmetry ($\gamma_4$) terms are included, the flat-bands acquire a significant dispersion and become overlapped with each other
( Fig.\,\ref{fig:band_trigonal}\,\textbf{a,b}). 
Theses bands can only be separated by applying a strong enough gate voltage between top and bottom layers (Fig.\,\ref{fig:band_trigonal}\,\textbf{c}). Using numerical simulations, we identify the parameter space of twist angle $\theta$ and applied voltage $U$ where the first conduction band is isolated (Fig.\,\ref{fig:single_particle}\,\textbf{a}). On the other hand, we find that there is barely any regime where the first valence band is isolated (Fig.\,\ref{fig:single_particle}\,\textbf{c}). Such a particle-hole asymmetry in the band structure is originated from $\gamma_4$ and $\Delta$ terms. The results are consistent with the experimental findings \cite{TDBGexp2019}, showing that the system at charge neutrality remains metallic unless a rather large vertical electric field is applied.  Furthermore, a correlated insulating phase is only observed on electron-doping side, consistent with the theoretically expected particle-hole asymmetry. 
Note that the bandwidth is not as flat as that of magic-angle TBG. However, the bandwidth is still small compared to the interaction scale which implies that strongly-correlated-physics can still arise.
Indeed, there is some debate regarding the bandwidth of magic angle TBG itself, with reported bandwidths ranging from 10-40 meV \cite{ChoiSTM2019}.

Another crucial difference compared to TBG is the absence of two-fold rotational symmetry, which protects  
the Dirac points in TBG. As a result,  
the physics of TDBG is controlled by a single narrow band (per spin per valley) rather than two as in TBG. 
The TDBG Hamiltonian has the following symmetries ($i$) three-fold rotation symmetry $C_3$, ($ii$) time-reversal symmetry $\T$, and, ($iii$) mirror reflection about the $x$-axis $M_y$ which only exists in the absence of vertical electric field, and ($iv$) SU(2) spin-rotation symmetry. Finally, we assume that in the small angle limit, there is valley charge-conservation symmetry $U(1)_\textrm{v}$, arising from the decoupling of Moir\'e and atomic lattice scale physics.

In addition, the conduction band within each valley carries a non-zero Chern number. 
In ordinary condensed matter systems, ${\cal T}$-symmetry forbids the existence of Chern bands. However, in Moir\'e systems, Chern bands carrying opposite Chern numbers for opposite valleys can arise due to the valley decoupling. The overall system still satisfies ${\cal T}$-symmetry which exchanges the two valleys. Therefore, spontaneous valley polarization would lead to a Chern band without explicitly breaking ${\cal T}$-symmetry \cite{Zhang2018, Ghaemi2012,Xie2018, Bultinck19}. 
At $U=0$, the reflection symmetry $M_y$ enforces $C=0$ for both valleys. At $U\neq 0$, the conduction band develops a non-vanishing Chern number computed numerically in Fig.\,\ref{fig:single_particle}\,\textbf{c} which is equal to $\pm 2$ for the parameter region corresponding to band isolation. The evolution of Chern number as a function of $U$ is further confirmed using symmetry indicator (Methods). \change{This can be also understood from the well-known behavior of a AB-stacked bilayer graphene under an electric field. Under the electric field, the bilayer graphene becomes gapped and accumulates opposite Berry curvatures at $\boldsymbol{K}_+$ and $\boldsymbol{K}_-$ valleys, which amounts to a Chern number $C_v=\pm 2$ for each valley. \cite{BLG_valley1, BLG_valley2, BLG_valley3, BLG_valley_exp}. 
}

Finally, we discuss the effect of applied magnetic field which influences the single-particle physics in two distinct ways. 
First, it couples to the electron spin via Zeeman effect leading to the splitting of bands with opposite spin by $2 \mu_B B$. 
Second, it couples to the electron orbital motion leading to modifications in the band structure. For out-of-plane field, the orbital effect arises from the magnetic field coupling to the planar motion of the electron \cite{Koshino2011, McEuen2017}. It leads to an energy correction of $\mu_B g^z_{\tau}(\bk) B_z$, with a $\bk$-dependent $g$-factor $g^z_{\tau}(\bk)$ satisfying $g^z_{-\tau}(-\bk) = -g^z_{\tau}(\bk)$ due to time-reversal symmetry ($\tau$ is a valley index). As shown in Fig.\,\ref{fig:single_particle}\,\textbf{f}, $g^z_\tau (\bk)$ can be much larger than the Zeeman effect. For in-plane field, the orbital effect arises from coupling to the interlayer motion of electrons. For an in-plane field $\bB$, we can choose the gauge $\bA(\bz) = - \bz \times \bB$ which does not depend on $x$ or $y$, thus preserving the Moir\'e translation symmetry. The resulting change in the hopping parameters is obtained by the Peierl's substitution, effectively providing an additional momentum shift of $-\frac{e}{\hbar} \frac{(l + m)d}{2}\,  \be_z \times \bB$ to the hopping connecting layers from $l$ to $m$, where $d$ is the interlayer separation (See Methods). This leads to an energy correction of the form $\mu_B (g^x_{\tau}(\bk) B_x + g^y_{\tau}(\bk) B_y)$ to the leading order in $\bB$ with $g^{x,y}_{-\tau}(-\bk) = -g^{x,y}_{\tau}(\bk)$.
The orbital effect due to in-plane field amounts to a very small relative momentum shift $\sim \frac{e d a}{\hbar} \approx 10^{-5}$.
However, it cannot be neglected since it is of the same order of magnitude as the Zeeman effect, $\frac{e v_F d}{\mu_B} \sim 1$ (see Fig.\,\ref{fig:single_particle}\,\textbf{d,e}). In general, the in-plane orbital contribution 
changes the band dispersion due to its $\bk$-dependence whereas the Zeeman effect shifts the entire band uniformly. Moreover, it acts oppositely for different valleys. These properties can be crucial in understanding the effect of in-plane field on the insulating gap and the superconducting temperature (See Methods and Supplementary Material 6).

\begin{figure*}[t]
\begin{center}
\includegraphics[width=0.95\textwidth]{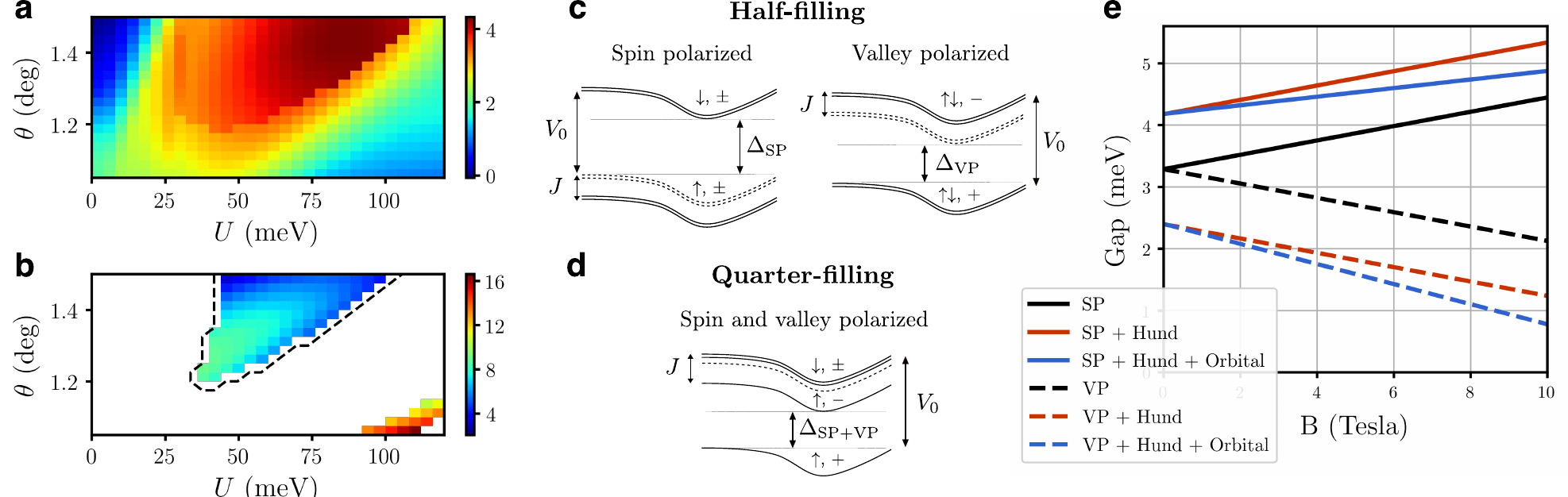}

\vspace{-0.1in}
\caption{ Results of the Hartree-Fock calculation. \textbf{a,} Color plot (meV) for $E_\textrm{IVC}  - E_\textrm{VP}$ per electron. \textbf{b,}  Color plot of self-consistency gap $\Delta_\textrm{SP/VP}$ for the SP/VP-state in the band isolated region. (No $J$-term included) \textbf{c,d } Effect of the intervalley Hund's coupling ($J$-term) on the gap for spin and valley polarized phases. at half and quater fillings, respectively. At half-filling, $J$-term increases (decreases) $\Delta_\textrm{SP}$ ($\Delta_\textrm{VP}$). At quarter-filling, $J$-term reduces the gap to the next-excited state, making the  quarter-filled insulator (SP+VP) less stable than the half-filled (SP) one. \textbf{e, } The correlated gap $\Delta$ for half-filling  insulators (SP,VP) as a function of in-plane $B_x$-field. $(\theta,U) = (1.33^\circ, 60 \textrm{ meV})$. Solid lines for SP-state and dotted lines for VP-state. Zeeman effect would increase (decrease) $\Delta$ for the SP (VP) state with increasing $B$. The valley orbital effect $g^{x,y}(\bk)$ leads to a linear decrease in the gap with field, thus effectively decreasing (increasing) the $g$-factor for the SPS (VP) state.}
\label{fig:HF}
\end{center}
\end{figure*}

\subsection{Correlated insulating states}
\label{sec:correlated}

In the band isolation regime, the first conduction band carries a non-zero Chern number as shown in Fig.\,\ref{fig:single_particle}\,\textbf{a,b} which prevents the existence of exponentially localized Wannier functions \cite{Vanderbilt1997}. As a result, one cannot construct a Hubbard model for the band unless valley-symmetry is broken or the model is enlarged to include more bands so that the net Chern number is zero. Instead of seeking a complicated real-space description, we discuss the interaction effects in the momentum space, as in the case of quantum-Hall-ferromagnetism. One major consequence of the absence of localized Wannier orbitals is the inadequacy of the Mott picture where the insulating phase is driven by strong repulsion between localized orbitals. Thus, we will use the terminology, \emph{correlated insulator} to refer to the interaction-driven insulating phase for the following physics.

In order to uncover the nature of the possible correlated insulating states at half and quarter filling \cite{TDBGexp2019}, we perform a self-consistent Hartree-Fock mean field theory similar to the one employed in Ref.~\cite{Po2018, Zhang2018}. Below, we sketch the derivation from the microscopic theory, relegating most details to Supplementary Material 2 and 3. The interacting Hamiltonian in momentum space is given by
\begin{equation}\label{eq:coulomb}
    {\cal H}_\textrm{int} = \frac{1}{2\, \textrm{Vol} } \sum_{\bq} \hat{\rho}(\bq) V(\bq) \hat{\rho}(-\bq),
\end{equation}
where $V(\bq)$ is the Fourier-transformed screened Coulomb interaction \cite{Young_FCI_2018, Screening_2}. Since the screening coming from the distance between the system and the gate is comparable to the Moir\'e length scale, the screening length can be important for the interaction effects. The density $\hat{\rho}(\bq)$ consists of an intravalley part $\rho^+ \sim c_\pm^\dagger c_\pm$ and an intervalley part $\rho^- \sim c_\pm^\dagger c_\mp$, where $c^\dagger_\pm$ is the electron creation operator for $\bK_\pm$-valley. The latter contribution arises from the small coupling between opposite valleys and gives rise to an intervalley Hund's coupling term.

The resulting Hamiltonian consists of two parts, $\H_{\rm int} = \H_{0} + \H_J$, where $\H_{0}$ contains the coupling between intravalley densities $\rho^+ \rho^+$  whereas $\H_J$ contains the coupling between intervalley densities $\rho^- \rho^-$. Rough estimation for the relative energy scales for $H_{0}$ and $H_J$ gives ${V_0} \sim 35 \textrm{ meV}$ and $J \sim 0.6 \textrm{ meV}$ for the experimentally relevant regime. Although $H_J$ is significantly smaller than $H_{0}$, it breaks the symmetry of the model down from two independent SU(2) spin-rotation symmetries for each valley to a single SU(2). Thus, it can lift the degeneracy between some symmetry breaking states which are degenerate on the level of the $H_0$. Indeed, we found that $H_J$ favors the spin alignment between opposite valleys and can be written in the form of inter-valley Hund's coupling as in \cite{Zhang2018}.

Within the self-consistent Hartree-Fock mean field theory, we consider the order parameter defined as
\beq
\langle c^\dagger_{\sigma, \tau}(\bk) c_{ \sigma',\tau'}(\bk') \rangle = M_{\sigma\tau, \sigma' \tau'}(\bk) \delta_{\bk,\bk'}.
\eeq
\change{For a gapped phase, matrix $M(\bk)$ must be a projector, i.e. $M(\bk)^2 = M(\bk)$ satisfying $\tr M(\bk) = \nu$ for all $\bk$.} Given that there are four flavors of fermions due to spin ($\sigma$) and valley ($\tau$) degeneracies, any possible order parameter $M$ can be expanded in terms of the generators of SU(4) $\sigma_i \otimes \tau_j$,
which can be grouped based on their symmetry breaking into 5 categories: 
\change{ (i) $\{ \sigma_0 \tau_z \}$ only breaks $\T$ and corresponds to a valley-polarized (VP) state, (ii) $\{ \sigma_{x,y,z} \tau_{0} \}$ breaks spin-rotation symmetry and correspond to a spin-polarized (SP) state. (iii) $\{ \sigma_{x,y,z} \tau_{z} \}$  breaks both spin-rotation and time-reversal (but preserve some combination of the two) and corresponds to a {spin-valley locked} (SVL) state, (iv) $\{\sigma_0 \tau_{x,y} \}$ breaks $U(1)$ valley-charge conservation and corresponds to an inter-valley coherent (IVC) state, and (v) $\{ \sigma_{x,y,z} \tau_{x,y}\}$ breaks both spin-rotation and U(1)$_v$ valley-charge conservation, corresponds to spin-IVC locked (SIVCL) state (see Table \ref{tab:orderparam}). We note that any of these orders may break or preserve $C_3$ symmetry depending on its $\bk$ dependence.
}

{\renewcommand{\arraystretch}{1.3}

{\renewcommand{\arraystretch}{1.3}

\begin{table}[t]
\begin{center}

\scalebox{0.88}{
\begin{tabular}{c||c|c}
$\nu=2$ & Example of the state &\, Sym. Gen. \\
\hline \hline
SP &\,  $\ket{\uparrow \boldsymbol{K}_+}\otimes \ket{\uparrow \boldsymbol{K}_-}$ \,& \, $U(1)_z$, $U(1)_v$, $\T$  \\
\hline
VP &\,  $\ket{\uparrow \boldsymbol{K}_+} \otimes \ket{\downarrow \boldsymbol{K}_+}$ \,& \, $SU(2)$, $\T$  \\
\hline
SVL &\,  $\ket{\uparrow \boldsymbol{K}_+} \otimes \ket{\downarrow \boldsymbol{K}_-}$ \,&\,  $U(1)_z$, $U(1)_v$, ${\T'}$  \\
\hline
IVC &\,  $\mqty{ \big( \ket{\uparrow \boldsymbol{K}_+} + e^{i\theta} \ket{\uparrow \boldsymbol{K}_-}\big) \otimes  \\ \big( \ket{\downarrow \boldsymbol{K}_+} + e^{i\theta} \ket{\downarrow \boldsymbol{K}_-} \big)}$ \,& $SU(2)$, $\T$  \\
\hline
\,SIVCL\, &\, $\mqty{ \big( \ket{\uparrow \boldsymbol{K}_+} + e^{i\theta} \ket{\downarrow \boldsymbol{K}_-}\big) \otimes  \\ \big( \ket{\downarrow \boldsymbol{K}_+} + e^{i\theta} \ket{\uparrow \boldsymbol{K}_-} \big)}$ \,& \, $U(1)_z$, $\mathbb{Z}_2^{xv}$, $ \T$  \vspace{0.12in}
\\

$\nu=1,3$ & Example of the state &\, Symmetry \\
\hline \hline
SVP &\,  $\ket{\uparrow \boldsymbol{K}_+} $ \,& \, $U(1)_z$, $U(1)_v$ \\
\hline
SPIVC &\,  $\ket{\uparrow \boldsymbol{K}_+} + e^{i\theta} \ket{\uparrow \boldsymbol{K}_-} $\, & $U(1)_z$, $\T$  \\
\hline
SVLIVC\, &\,  $\ket{\uparrow \boldsymbol{K}_+} + e^{i\theta} \ket{\downarrow \boldsymbol{K}_-} $\, &\, $\mathbb{Z}_2^{zv}$, $\T'$  \\
\end{tabular}}

\caption{\label{tab:orderparam}
\change{ Examples of symmetry broken states and corresponding remaining symmetries for all possible translation-symmetric gapped states at half $\nu=2$ and quarter $\nu=1$ fillings. The similar table with the form of $M(\bk)$ and symmetry generators is in Supplementary material 3. Here, $\T$ is the spinless time-reversal $\T=\tau_x \K$ squaring to $+1$ whereas $\T'$ is the spinful time-reversal $\T'= i \sigma_y \T$ squaring to $-1$ (with $\K$ denoting complex conjugation).  $U(1)_{x,y,z}^\theta = e^{i \theta \sigma_{x,y,z}/2}$ denotes spin rotation around the $x, y, z$ axis by an angle $\theta$ whereas $U(1)_v^\theta = e^{i \theta \tau_z/2}$ denotes rotation in the valley $x-y$ plane by an angle $\theta$. Finally, $\mathbb{Z}_2^{z, v}$ is generated by the combined rotation $U(1)_{z}^{\pi} U(1)_v^{\pi}$}.  }
\end{center}
\end{table}
}

The results of the self-consistent Hartree-Fock calculation are summarized in the following (See Supplementary Material 3 for details). \change{Restricting ourselves to translation-symmetric gapped states, we find there are five options: SP, VP, SVL, IVC, and SIVCL at half-filling $\nu = 2$ and three options: spin-valley-polarized (SVP), spin-polarized-IVC (SPIVC), and spin-valley-locked-IVC (SVLIVC) at quarter-filling $\nu = 1,3$, as in Tab.\,\ref{tab:orderparam}}. By solving the Hartree-Fock self-consistency condition, the ground state energy $E$ and the correlation gap $\Delta$ are computed for different states,  Fig.\,\ref{fig:HF}\,\textbf{a}. \change{Let us first consider what happens in the absence of intervalley Hund's coupling. In this case, we find that the SP and SVL states at half-filling and similarly the SPIVC and SVLIVC states at quarter-filling are exactly degenerate since they are related by a spin-rotation in one of the valleys. Similarly, due to the enlarged symmetry of the mean-field Hamiltonian, the SP and VP states and the IVC and SIVCL states have the same energy (see supplemental material). Thus, we only need to numerically investigate the competition between SP and IVC at half-filling and SVP and SPIVC at quarter-filling. The result of such numerical investigation is shown in Fig.~\ref{fig:HF}\,\textbf{a} where we clearly see that SP has a lower energy than that of the IVC in most of the parameter regime. Similar results apply for the competition between SVP and SPIVC at quarter-filling. 
The correlation-induced gap $\Delta$ for the SP state in the band isolation region ranges between 4 and 8 meV, see Fig.\,\ref{fig:HF}\,\textbf{b}. 

To understand the reason why IVC order is energetically unfavorable, we can employ the argument of Ref.~\cite{Bultinck19} as follows. IVC order between two valleys with opposite Chern number $C$ is equivalent after a particle-hole transformation in one of the valleys to superconducting pairing between bands with the \emph{same} Chern number i.e. a superconductor in a background magnetic field. This means that the order parameter necessarily includes $|C|$ vortices within the Brillouin zone leading to increased energy. A more detailed analytic treatment of the energy competition between SP and IVC is provided in the supplemental material.}


The inclusion of the effect of intervalley Hund's coupling alters the competition between the phases as follows. First, \change{ since the term is ferromagnetic}, it lowers the energy of the SP-state, favoring the SP-state over the VP-state \change{ which is in turn favored over the SVL-state}. Second, it lowers the energy of the filled bands for the SP-state at half-filling, thus increasing $\Delta_{\rm SP}$. On the other hand, it reduces the energy of some of the empty bands for the VP-state, reducing $\Delta_{\rm VP}$ (see Fig.\,\ref{fig:HF}\,\textbf{c,e}). 
The Hund's coupling term similarly reduces $\Delta_{\rm SVP}$ at quarter filling by lowering the energy of one of the excited states (see Fig.\,\ref{fig:HF}\,\textbf{d}).  We note here that the reduction of the correlated gap at quarter filling relative to that at half-filling may explain why the former is more difficult to observe experimentally compared to the latter and requires the application of a magnetic field \cite{TDBGexp2019}. 

In the presence of an in-plane field, the gap of the SP-phase at half-filling is expected to grow with a slope consistent with the Zeeman $g=2$ factor. However, the orbital effect discussed earlier leads to a reduction in the effective $g$-factor by 20-50\% depending on the band structure details (Fig.\,\ref{fig:HF}\,\textbf{e}), which is in agreement with the experimental data \cite{TDBGexp2019}. From the numerical calculation, we confirmed that such a reduction in gap also depends on the in-plane field direction, which exhibits 3-fold periodicity (see Methods). Therefore, the orbital effect can be directly verified in a rotating in-plane field setup, where we predict the modulation of the $g$-factor with period $2\pi/3$ in the angle.

\begin{figure}[t]
\begin{center}
\includegraphics[width=0.44\textwidth]{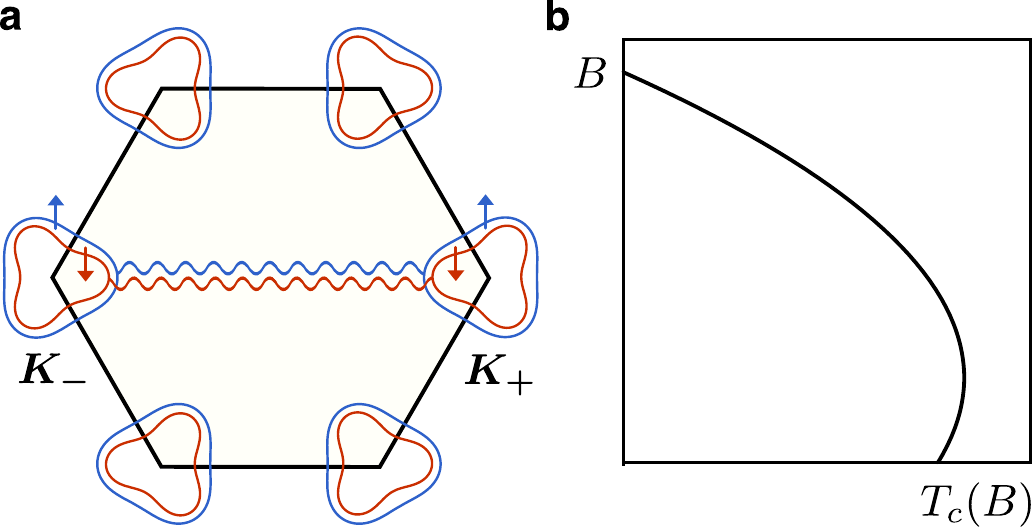}

\vspace{-0.1in}
\caption{ \textbf{a, } Superconductor triplet paring between the $c_{\sigma,+}(\bk)$ and $c_{\sigma,-}(-\bk)$ with exact energy match. \textbf{b,} Schematic plot for the $T_c$ as a function of $B$-field.  }
\label{fig:SC}
\end{center}
\end{figure}

\subsection{Superconductivity}
\label{sec:superconductivity}
When the correlated insulator is doped away from half-filling, a superconducting phase is observed below 3.5 K \cite{TDBGexp2019}. Our proposed scenario for the observed superconductivity is illustrated in Fig.\,\ref{fig:SC}\,$\mathbf{a}$ where pairing takes place between time-reversal partners in opposite valley. \change{Such an intervalley pairing between time-reversal partners has also been proposed \cite{SC_TMD_theory1, SC_TMD_theory2, SC_TMD_theory3} and observed in transition metal dichalcogenides (TMD) \cite{SC_TMD_exp}. However, unlike in TMD, where strong spin orbit coupling implies a locking between spin and valley, here the proposed pairing takes place between the electrons with the same spin. }  To understand this, we first note that doping a spin-polarized insulator is expected to give rise to a ferromagnetic metal with spin-split Fermi surface. Similar to other ferromagnetic metals \cite{saxena_superconductivity_2000, aoki_coexistence_2001, huy_superconductivity_2007}, ferromagnetic spin fluctuations can act as a pairing glue responsible for superconductivity  \cite{TBG_SC_triplet2018}. This motivates the following simplified Hamiltonian,
\beq
\label{HF}
\H = \sum_{\bk, \tau, \sigma} c_{\sigma, \tau, \bk}^\dagger \xi_{\sigma, \tau, \bk} c_{\sigma, \tau, \bk} - g \sum_\bq \bS_\bq \cdot \bS_{-\bq},
\eeq
where the spin operator $\bS^a_\bq = \sum_{\bk, \tau, \sigma, \sigma'} c^\dagger_{\sigma, \tau, \bk + \bq} \bsigma^a_{\sigma, \sigma'} c_{\sigma', \tau, \bk}$. This Hamiltonian can be obtained within an RPA treatment by identifying the ferromagnetic order as the leading instability in the doped itinerant phase. The ferromagnetic susceptibility is peaked at $\bq = 0$ which justifies a $\bk$-independent coupling.

Next, we consider the simplest possible intervalley superconducting pairing function $\Delta$ which is $\bk$-independent ($s$-wave) within each valley. Note, however, that the overall orbital symmetry incorporating both momentum and valley may still be anti-symmetric, e.g. $p$-wave. For the proposed pairing, $\Delta$ is proportional to $\tau_x$ or $\tau_y$ corresponding to valley triplet or singlet, respectively. The overall antisymmetry of $\Delta$ implies that the former scenario corresponds to a spin-singlet $i \sigma_y$ whereas the latter corresponds to a spin-triplet $i \sigma_y \bd \cdot \bsigma$. Here, $\bd$ is the vector which captures the direction of the spin state. \change{To see which of these is the dominant pairing channel, it is useful to decouple the interaction in the pairing channel as
\beq
\H_{\rm int} = -g \sum_{\bk,\bq} \tr (\bsigma \Delta_\bk) \cdot (\bsigma^T \Delta_{\bk+\bq}^\dagger )
\eeq
We now assume $\bk$-independent $\Delta$ and decompose it into spin-singlet/velly triplet $\Delta_s$ and spin-triplet/valley singlet $\Delta_t$. We now use
\beq
\label{Gap}
 \bsigma \cdot (\Delta_{t,s} \bsigma^T) = \lambda_{t,s} \Delta_{t,s}, 
 \eeq
 where $\lambda_t = 1$ and $\lambda_s = -3$. This means that the interaction is repuslive in the singlet channel and attractive in the triplet channel making the latter the dominant pairing channel. A more detailed discussion of these pairing channels within the linearized BCS equation is provided in Supplementary Material. } 

We highlight here that spin-triplet pairing is only known to occur in liquid He${}_3$ \cite{Ambegaokar73} and a few Uranium compounds \cite{saxena_superconductivity_2000, aoki_coexistence_2001, huy_superconductivity_2007} as it requires   pairing that varies over the Fermi surface (eg. $p$-wave) which is likely to be energetically unfavorable in typical solids. The existence of the {\em valley} degree of freedom here enables us to evade this difficulty and obtain a spin-triplet valley-singlet order parameter even for a $\bk$-independent interaction.

The experimental consequences of the proposed spin-triplet valley-singlet superconductivity can be investigated by writing the Ginzburg-Landau free energy for the order parameter $\Delta = \tau_y \sigma_y \bd \cdot \bsigma$ in the presence of a magnetic field $\bB$. Restricting ourselves to terms up to quartic order in $\bd$ or $\bB$, we can write the following free energy functional 
\change{
\begin{multline}
\label{GLF}
F = \kappa \left[(T - T_c + b (\mu_B \bB)^2) \bd \cdot \bd^* + i a \mu_B \bB \cdot (\bd \times \bd^*) \right. \\ \left. + c \mu_B^2 |\bB \cdot \bd|^2 + \alpha (\bd \cdot \bd^*)^4 + \eta |\bd \cdot \bd|^4\right]
\end{multline}
Detailed microscopic derivation of the coefficients $a, b, c, \kappa, \alpha, \eta$ is provided in supplemental material.} In the absence of spin-orbit coupling, the order parameter's spin is expected to align with the magnetic field. \change{Assuming the magnetic field is parallel to the $z$-axis, $\bB = B \be_z$, we can then write
\beq
\bd = (\frac{\Delta_{\uparrow \uparrow} + \Delta_{\downarrow \downarrow}}{2}, \frac{\Delta_{\uparrow \uparrow} - \Delta_{\downarrow \downarrow}}{2i}, 0)
\eeq
Substituting in the free energy (\ref{GLF}) and using the fact that $\eta = -\alpha/2$ (see supplemental material) yields
\begin{gather}
F = \frac{\kappa}{2} \sum_{s=\uparrow,\downarrow} F_s \nonumber \\ F_s =  |\Delta_{ss}|^2(T - T_c - \sigma_s a \mu_B B + b (\mu_B B)^2)  + \frac{\alpha}{2} |\Delta_{ss}|^4
\label{GLFDelta}
\end{gather}
One important feature is that $\alpha > 0$ which implies the stability of the phase considered.}

The free energy (\ref{GLFDelta}) leads to the following dependence of the superconducting $T_c$ on the applied field
\beq
\label{TcB}
T_{c,\uparrow/\downarrow}(B) = T_c \pm a \mu_B B - b (\mu_B B)^2.
\eeq
The most remarkable feature of this result is that, for non-zero $a$, $T_c$ initially increases upon the application of magnetic field. This can be understood as follows: for a ferromagnetic metal with weakly spin-split Fermi surfaces, the application of the Zeeman field increases (decreases) the density of states for the majority (minority) spin Fermi surface, leading to a linear increase in $T_c$ for the majority spin with the coefficient
\beq
a = 2\chi T_c \frac{N'(0)}{N(0)} \ln \frac{\Lambda}{T_c}
\label{aB}
\eeq
where $\Lambda$ is the bandwidth, $N(0)$ is the density of states at the Fermi energy, and $\chi$ is the dimensionless magnetic susceptibility (see supplemental material). Similar linear field-dependence of $T_c$ is known in superfluid He${}_3$ \cite{Ambegaokar73}, indicating independent pairing for each spin species. This behavior is in stark contrast to the monotonic decrease of $T_c$ under increasing $B$-field in a spin-singlet superconductor. 
One crucial observation here is that $a$ seems to depend on several details and is expected to be very small since $T_c \ll \frac{N(0)}{N'(0)} \sim \epsilon_F$. Surprisingly, the measured value of $a$ is of order 1 \cite{TDBGexp2019} which suggests the vicinity of a quantum critical point where the scaling of the susceptibility cancels exactly against the other parameters. Indeed, the scaling $\chi \sim \epsilon_F/(T \log T)$ predicted by Herz-Millis theory in the quantum critical regime for an itinerant ferromagnet \cite{Herz76, Millis93} leads to such cancellation resulting in $a \sim 1$.

The origin of the quadratic term in Eq.~\ref{TcB} can be understood in terms of the in-plane orbital effect discussed in Sec.~\ref{sec:single_particle}. First, note that Zeeman splitting cannot break Cooper pairs between aligned spins. Instead, it yields an initial linear increase in $T_c(\bB)$ followed by saturation at large fields when all the spins are aligned. On the other hand, the in-plane orbital effect can induce pair-breaking by mismatching the energies of time-reversal partner states in opposite valleys, resulting in a quadratic decrease in $T_c$ with the applied field whose coefficient is given by (see Supplementary Material 6)
\beq
b= \frac{1}{T_c} \int_{\rm FS} d\bk  (\be_{\bB} \cdot \bg_{+,\bk})^2
\eeq
where $\be_{\bB}$ is the direction of the external magnetic field. The average value of $(\be_\bB \cdot \bg_+(\bk))^2$ over the Fermi surface depends strongly on the filling and the field direction with typical value around 1 (cf.~Fig.~\ref{fig:single_particle}d-f). Using this value, we can make a rough estimate for the in-plane field needed to destroy superconductivity as $\mu_B B_c \sim \sqrt{T_c/b}$ yielding a value about 3 Teslas which compares favorably to the experimental value \cite{TDBGexp2019}.
Furthermore, if we consider an out-of-plane field instead, $|g_z|$ is on average about 1-2 orders of magnitude larger than $|g_{x,y}|$, yielding a critical field of about $\sim 0.1 T$ which is very close to the experimentally observed result \cite{TDBGexp2019}. 

It is worth noting that the reduction of $T_c$ at large field can also arise from the suppression of ferromagnetic fluctuations responsible for the pairing, as has been observed in the ferromagnetic superconductor UCoGe \cite{Hattori11}. Such effects are neglected within our simplified analysis (\ref{HF}) which assumes a constant coupling $g$.

\section{Conclusion}

\change{
In this work, we theoretically investigated the physics of twisted double bilayer graphene (TDBG), addressing the experimental observations of correlated insulating phases at integer fillings and the neighboring  superconductor reported in Ref.~\cite{TDBGexp2019}. 

First, let us summarize a few important features of the band structure. Due to the absence of a $C_2$ symmetry in TDBG, isolated conduction and valence bands with non-zero valley Chern numbers can exist. Moreover, trigonal warping and particle-hole asymmetry in each bilayer graphene lead to (i) a significant broadening of each band so that they overlap in the absence of a displacement field, and (ii) asymmetry between electron- and hole- doped systems. As a result, the parameter space that can host strongly correlated physics is significantly constrained, and the tunability from displacement field at a particular filling becomes essential to realizing correlated states.

Second, we identified an important role played by the coupling of in-plane field to the orbital motion of the electron in TDBG. Despite being small compared to the bandwidth, this effect is comparable to Zeeman splitting, leading to a modified $g$-factor which compares favorably to the experimental value \cite{TDBGexp2019} extracted from the slope of the half-filling gap as a function of in-plane field. Moreover, in our theory, this effect is responsible for the reduction of $T_c$ under an in-plane field by providing the main pair breaking mechanism when pairing takes place between aligned spins in opposite valleys. The resulting decrease in the superconducting $T_c$ with in-plane field agrees qualitatively with the experimental results.}

\change{Furthermore, we have performed a self-consistent Hartree-Fock mean field calculation to identify the possible symmetry broken correlated insulating states at integer fillings. Our prediction of a spin-polarized ferromagnet at half-filling is consistent with the observed increase in the gap with in-plane field.

Finally, here we have proposed a pairing mechanism based on ferromagnetic fluctuations, which is motivated by the evidence for a ferromagnetic parent insulator. Such a mechanism leads naturally to the spin-triplet pairing suggested by experiments. In addition, we showed that the experimentally observed dependence of $T_c$ on in-plane field  suggests that the superconductor emerges in the vicinity to a quantum critical point. 
}

In conclusion, our theoretically established phase diagram for twisted double bilayer graphene,  captures all significant observations of the experiments reported in   \cite{TDBGexp2019}. This includes  single-particle features such as the parameter range for band isolation as well as correlation-induced features including a ferromagnetic insulator at half-filling which leads to a spin-triplet superconductor upon doping. In addition to deepening our understanding of correlated Moir\'e materials, our results highlight how phases which are rare in conventional solids can be readily realized in this novel and tunable platform. 

\vspace{5pt}

\noindent {\bf Note}:  After completing this work we noticed two experimental papers  \cite{PabloTDBG,IOP_TDBG}  which are consistent with Ref.~\cite{TDBGexp2019} and theoretical discussion contained here.

\section{Methods}
{\small 
\noindent {\bf Numerical Simulations for Single Particle} Here, we summarize the numerical methods used to calculate the single particle physics. First, each bilayer graphene (BLG) layer is modeled by the following bloch Hamiltonian:
\beq
\label{eq:blg}
h_\bk=\mqty( U_{1}+\Delta & - \gamma_0 f(\bk) & \gamma_4 f^*(\bk) & \gamma_1 \,\, \\
		 - \gamma_0 f^*(\bk) & U_{1} & \gamma_3 f(\bk) & \gamma_4 f^*(\bk) \\
        \gamma_4 f(\bk) & \gamma_3 f^*(\bk) & U_{2} & -\gamma_0 f(\bk) \\
        \gamma_1& \gamma_4 f(\bk) & - \gamma_0 f^*(\bk) & U_{2}+\Delta ),
\eeq
which is labelled in the order of  $A_\textrm{1}$, $B_\textrm{1}$, $A_\textrm{2}$, $B_\textrm{2}$.
Here, we consider a realistic model of BLG illustrated in Fig.~\ref{fig:ABAB_schematic}.  AB-stacking means that the $A$-site of the first layer ($A_1$) sits on top of the $B$-site of the second layer ($B_2$). This gives a small on-site energy $\Delta$ for these sites. 
Here, $f(\bk) \equiv \sum_l e^{-i \bk \cdot \delta_l}$,
where $\delta_1 = a(0,-1)$, $\delta_2 = a(-\sqrt{3}/2,1/2)$, and $\delta_3 = a(\sqrt{3}/2,1/2)$ are vectors from $B$-site to $A$-sites.
One can expand $f(\bk)$ near $\bK_\pm = \pm(4\pi/3\sqrt{3}a,0)$ as
\beq
f(\bK_\pm + \bk) = \frac{3}{2} (\mp k_x + ik_y) a,
\eeq
where $a$ is the distance between carbon atoms. 
Throughout, we will use the phenomenological parameters extracted from  Ref.~\cite{Jeil2014}
\beq\label{eq:param}
(\gamma_0, \gamma_1, \gamma_3, \gamma_4, \Delta) = (2610, 361, 283, 138, 15) \textrm{ meV}.
\eeq
where $\gamma_{0,1,3,4}$ and $\Delta$ are the parameters illustrated in Fig.~\ref{fig:ABAB_schematic}. Additionally, the potential difference between the top and bottom graphene layer, $U$ is an important parameter in the experiment, which is controlled by the gate voltage difference. For a displacement field strength $D$, ABAB system's dielectric constant $\epsilon$ and the thickness of the BLG/BLG system $d$, $U = {\epsilon^{-1}}D\cdot d$.

Next, we couple two layers of AB-stacked bilayer graphenes by Moire hoping terms. As we are interested in the physics near charge neutrality point, we focus on band structures mostly originated near $\bK_\pm$ points. In the continuum model approximation \cite{MacDonald2011}, Moire bands from $\bK_\pm$ valleys decouple; for the Moire band from $\bK_+$ valley, the Hamiltonian is given by  
\begin{align}\label{eq:HABAB}
    H_+ &= \sum_\bk \Big[\, h^t_{\frac{\theta}{2} }(\bK_+ + \bk) c_{\bk,+,t}^\dagger c^{}_{\bk,+,t}  + h^b_{-\frac{\theta}{2} }(\bK_+ +\bk) c_{\bk,+,b}^\dagger c^{}_{\bk,+,b} \nonumber \\
    & + \sum_n \qty( T_n c^\dagger_{\bk+\bq_n,+,b} c^{~}_{\bk,+,t} + T^\dagger_n c^\dagger_{\bk,+,t} c^{~}_{\bk+\bq_n,+,b} ) \Big], 
\end{align}
where $c^\dagger_{\bk,+,t/b}$ is a 4-components electron creation operator for top/bottom layer with momentum $\bK_+ + \bk$. Here, $h_\theta(\bk) = h(R_{-\theta} \bk)$ with $R_\theta$ denoting the counter-clockwise rotation matrix by angle $\theta$ relative to the $x$-axis. The momenta $\bq_{0,1,2}$ are given by $\bq_0 = R_{\theta/2} K - R_{-\theta/2} K = \frac{8\pi \sin (\theta/2)}{3\sqrt{3} a} (0, -1)$, $\bq_1 = R_{\phi} \bq_0$, and $\bq_2 = R_{-\phi} \bq_0$ where $\phi = 2\pi/3$. The hopping matrices $T_n$, $n=0,1,2$ are given by
\begin{equation}
    T_n = \mqty(0&1\\0&0)_\textrm{layer} \otimes \qty(w_0 + w_1 e^{2\pi n \sigma_3 /3} \sigma_1 e^{-2\pi n \sigma_3 /3})_\textrm{sublattice},
\end{equation}
where $w_0,w_1$ are Moir\'e hopping parameters. 
One crucial parameter tunable in experiments is displacement field $U$. In Fig.\,\ref{fig:band_evolution}, we demonstrated how the band structure evolves with increasing $U$. One can see that the first conduction band becomes isolated in the range of $U \in [40,80]$. 
Furthermore, to illustrate the how the band isolation arises, we plot the energy gap between different bands in Fig.\,\ref{fig:gaps_bands}. For a smaller value of $r$, gapped regimes in Fig.\,\ref{fig:gaps_bands}\,\textbf{a,b,c}  expand in the parameter space of $(\theta,U)$, giving arise to a wider band isolation regime (Data available upon request).


\begin{figure}[t]
\begin{center}
\includegraphics[width=0.99\columnwidth]{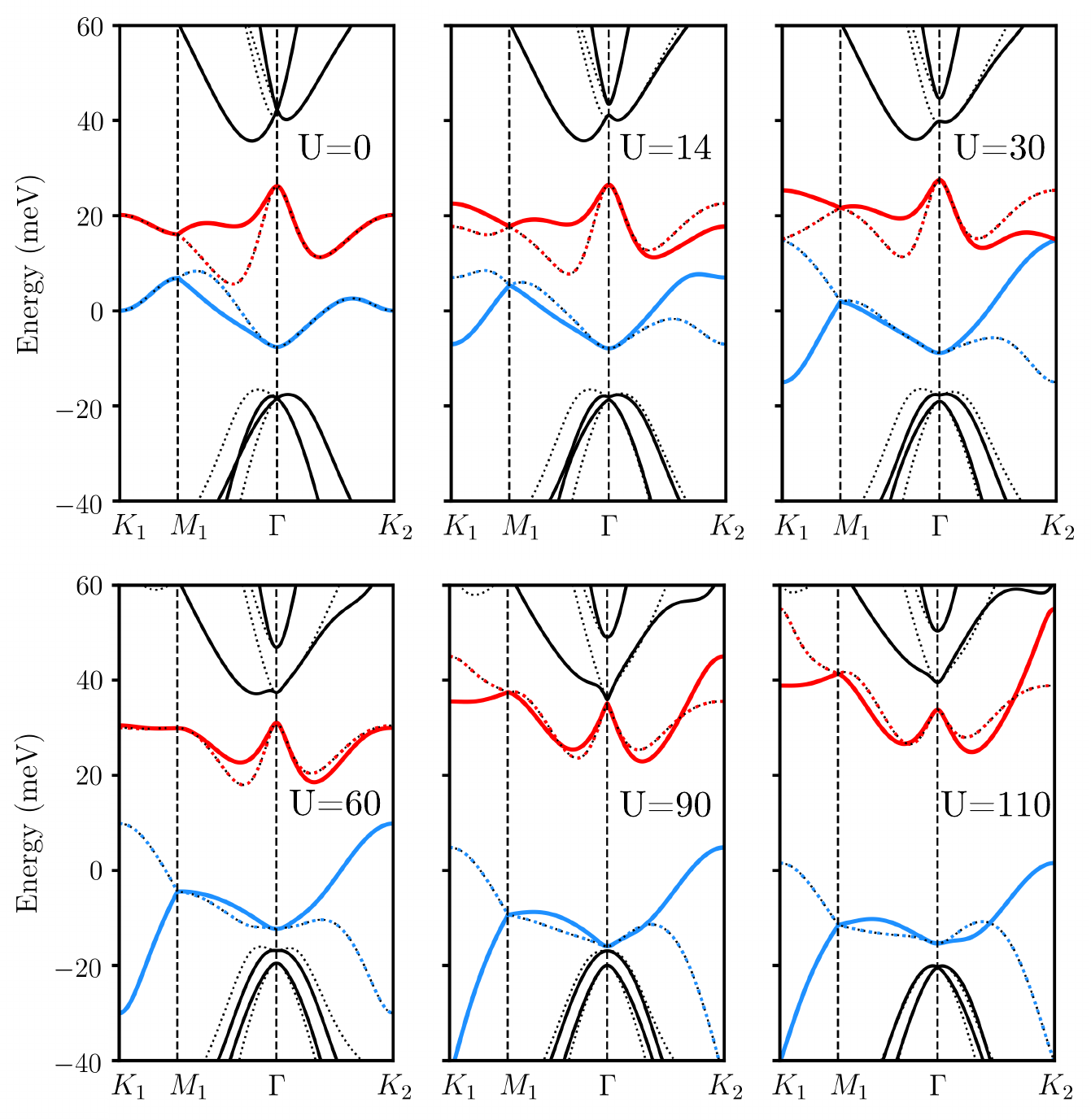}
\caption{ The band structure of the model at $\theta = 1.33^\circ$ and $U = 0,14,30,60,90,110$. At $U=14$, Chern number is exchanged by $3$ between the conduction and valence band at three momenta which are located not along the symmetric cut. However, at $U=30$ and $U=90$, Chern number changes by $1$ which can be seen by the gap closing between bands at $K_2$ and $\Gamma$ points.  }
\label{fig:band_evolution}
\end{center}
\end{figure}

\begin{figure}[t]
\begin{center}
\includegraphics[width=0.95\columnwidth]{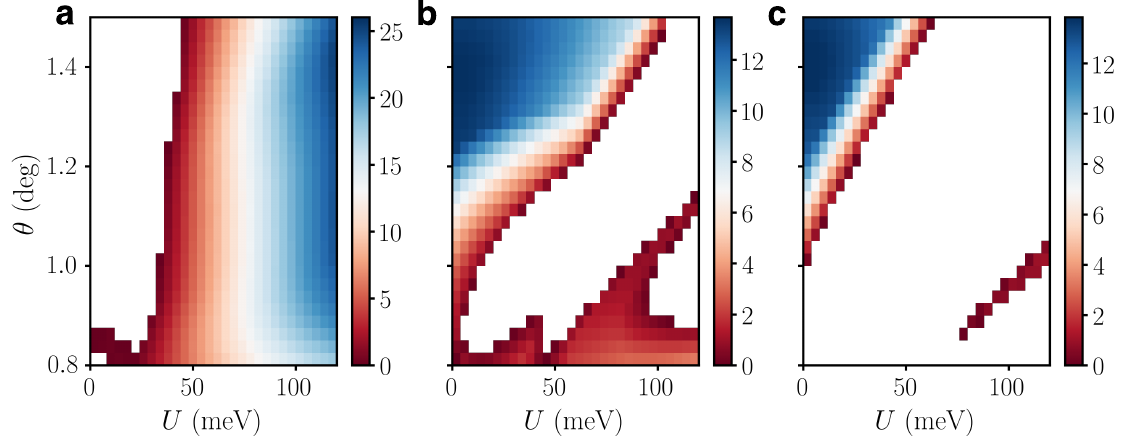}

\caption{The color represents bandgap in meV unit for the range of $(\theta,U)$. Uncolored region implies bands being overlapped. \textbf{a,} Gap between the first conduction and valence bands. \textbf{b,} Gap between the first and second conduction bands. \textbf{c,} Gap between the first and second valence bands.}
\label{fig:gaps_bands}
\end{center}
\end{figure}

\vspace{10pt} 
\noindent {\bf Chern Number} In the main text, we presented Chern number carried by Moire first conduction bands from $\bK_\pm$-valleys. Here, we carefully examine the evolution of Chern. First, at $U=0$, the reflection symmetry $M_y$ enforces $C=0$ for both valleys as $M_y$ maps the system back to itself without exchanging valleys, but $k_y \mapsto -k_y$ so Berry curvature flips its sign \cite{Zhang2018}. In the quadratic band approximation limit of BLG, as we increase $U$, the band inversion between conduction and valence bands occurs at the Moir\'e $K_2$-point ($K_1$ for negative $U$) with a quadratic touching. Thus, Chern number of $\pm 2$ is exchanged.

Next, let us understand the Chern number evolution in the realistic Hamiltonian with parameters of Eq.~\ref{eq:param} along the dotted line in Fig.\,\ref{fig:single_particle}\,\textbf{b}. With a trigonal warping term, the quadratic band touching point splits into four Dirac cones, three with positive and the other with negative chirality. These three Dirac cones are located at generic momenta, thus would not be observed in the band plot along the high symmetric line.
Under the presence of particle-hole asymmetry terms, the degeneracy between four Dirac cones split, and the band inversion would happen first at three Dirac cones, exchanging Chern number by $\pm 3$. Then, the band inversion would occur at the center Dirac cone, exchanging Chern number by $\mp 1$. In total, it will still change the Chern number by $\pm 2$. At larger values of the gate voltage $U$, the band inversion happens between first and second conduction band at $\Gamma$ point, and the Chern number then changes by $\mp 1$ (It can change by $\mp 2$ for other parameter setting), decreasing the Chern number.

This can be further checked by inspecting symmetry indicators \cite{Bernevig2012, Turner2012, Watanabe2017}. There are three $C_3$-invariant momenta $\Gamma$, $K$, and $K'$. For a Bloch state with these momenta, $C_3$ rotation symmetry would map the state back to itself with a rotation eigenvalue:
\begin{equation}
    R_{2\pi/3} \ket{\bk,n} = e^{2 \pi i L_{n,\bk}/3} \ket{\bk,n}, \quad \bk = K_1, K_2, \Gamma
\end{equation}
where $L_{n,\bk}$ is an angular momentum associated with the Bloch state $\ket{\bk,n}$.
Then, the Chern number of the $n$-th band can be determined modulo 3 by
\beq
C_n \equiv  L_{n,\Gamma} + L_{n,K_1} + L_{n,K_2} \mod 3 
\eeq
Thus, by tracking how $C_3$ eigenvalues of the three momenta change with the gating voltage $U$, we can understand how Chern number transition happens in the system. Indeed, the aforementioned scenario can be confirmed. For example, consider a Moir\'e first conduction band for $\bK_+$ valley at $\theta = 1.33^\circ$. At $U=0$ meV, we start with $(n_\Gamma, n_{K_1}, n_{K_2}) = (0,1,-1)$. At $U=14$ meV, Chern number changes by $+3$ but it can be only captured by Berry curvature not by symmetry indicator. At $U=30$ meV, Chern number changes by $-1$, manifested by $n_{K_2}: -1 \mapsto 1$. At $U=90$ meV, Chern number again changes by $-1$, manifested by $n_{\Gamma}: 0 \mapsto -1$.
See Fig.~\ref{fig:band_evolution} for the detail.

\vspace{10pt}

\noindent {\bf Magnetic Field Effect}
Under in-plane magnetic field $\bB = (B_x, B_y, 0)$, one can choose the gauge $\bA(\bz) = - \bz \times \bB$. Then, the effect of a magnetic field on hopping terms is evaluated via Peierl's substitution, where the hopping term from $\bR$ to $\bR+\bdelta$ is multiplied by the phase factor
\begin{equation}
    e^{i\frac{q}{\hbar} \int_{\bR}^{\bR+\bdelta} d \br \cdot \bA(\bz)} = e^{ -i\frac{e}{\hbar} \bdelta_{xy} \cdot \left[\qty(\bR_z + \frac{\bdelta_{z}}{2}) \times \bB \, \right]},
\end{equation}
such that 
\begin{equation}
    \sum_{\bR,\bdelta} e^{i\frac{q}{\hbar} \int_{\bR}^{\bR+\bdelta} d \br \cdot \bA(\bz)} c^\dagger_{\bR+\delta} c_{\bR} = \sum_{\bk,\bdelta} e^{-i \qty( \bk + \boldsymbol{\alpha} )\cdot \bdelta }  c^\dagger_\bk c_\bk,
\end{equation}
where $\boldsymbol{\alpha} = -\frac{q}{\hbar} \bA( \bR_z + \bdelta_z/2) = - \frac{e}{\hbar} \left[\qty(\bR_z + \frac{\bdelta_{z}}{2}) \times \bB \, \right]$ since $\bA(\bz)$ is linear function of $\bz$.  
Hence, the effect of in-plane field can be included by simply replacing all $\bk$-dependent matrix elements of Bloch Hamiltonians by $\bk + \boldsymbol{\alpha}$ as follows (we take $c_\bk = \sum_\bR e^{-i \bk \cdot \bR} c_\bR$):
\beq
\H_{l,m}(\bk, \bB) = \H_{l,m}\left(\bk - \frac{e}{\hbar} \frac{(l + m)d}{2}\,  \be_z \times \bB\right)
\eeq
where $\H_{l,m}$ is the matrix element connecting layers $l$ and $m$ ($l,m = 0,\dots,3$ from bottom to top) in Eq.\,\ref{eq:HABAB}, $d=3.42 \textrm{\AA}$ is the interlayer distance, and $\be_z$ is the unit vector in the $z$ direction.

Due to its small magnitude relative to the energy gap, it suffices to consider the in-plane orbital effect to first order in pertrubation theory. This amounts to adding the following in-plane orbital term to the single particle energies
\beq
\xi_{n,\tau}(\bk, \bB) = \xi_{n,\tau}(\bk) + \mu_B \bg^{xy}_{n,\tau}(\bk) \cdot \bB 
\eeq
where $\bg^{xy}_{n,\tau}(\bk)$ is given by
\beq
\bg^{xy}_{n,\tau}(\bk) = \frac{1}{\mu_B} \langle \psi_{n,\tau}(\bk)| \nabla_\bB \H_\tau(\bk,\bB)|_{\bB=0} |\psi_{n,\tau}(\bk) \rangle
\label{gxy}
\eeq
where $\tau$ is the valley index. Time-reversal symmetry implies that $\bg^{xy}_{n,\tau}(-\bk) = -\bg^{xy}_{n,-\tau}(\bk)$. The in-plane orbital $g$-factor transforms under $C_3$ rotation as
\beq
\bg^{xy}_{n,\tau}(R_{\pm 2\pi/3} \bk) = R_{\mp 2\pi/3} \bg^{xy}_{n,\tau}(\bk) 
\eeq
provided that the band $n$ is non-degenerate at $\bk$. This implies that $\bg^{xy}_{n,\tau}(\bk)$ vanishes at any $C_3$-invariant point. As pointed out in the Results, in general, the in-plane orbital contributions affects the bands very differently from the Zeeman effect. For example, it can distort the Fermi surface when the bands are partially filled in an opposite way in the two valleys which can influence the physical properties, e.g. superconducting $T_c$ (See Supplementary Material 6).

The effect of out-of-plane field on the energy bands is generally more complicated since any gauge choice breaks translation symmetry. As a result, the band picture breaks down for large enough out-of-plane fields where Landau level physics form instead. In the following, we will consider the limit of weak out-of-plane fields which can be treated perturbatively. In this case, the out-of-plane field induces an orbital valley Zeeman effect as pointed out in Ref.~\cite{Koshino2011, McEuen2017} whose $g$-factor is given by
\beq
 g^z_{n,\tau}(\bk) = -\frac{4 m}{\hbar^2} \Im \sum_{l\neq n} \frac{\langle n,\tau |\partial_{k_x} \H_\tau| l,\tau\rangle \langle l,\tau, |\partial_{k_y} \H_\tau| n\rangle}{\epsilon_{n,\tau,\bk} - \epsilon_{l,\tau,\bk}}
 \label{gz}
\eeq
In summary, the single particle energies has the following dependence on magnetic field
\beq
\xi_{n, \bsigma, \tau}(\bk, \bB) = \xi_{n, \tau}(\bk) + \mu_B \qty( g\bsigma \cdot \bB + \bg_{n,\tau}(\bk) \cdot \bB ),
\eeq
where $\bsigma$ is the electron spin operator (which is $\pm 1/2$ for up/down spins) and $\tau = \pm$. The valley orbital $g$-factor is defined as
\beq
\bg_{n,\tau}(\bk) = (\bg^{xy}_{n,\tau}(\bk), g^z_{n,\tau}(\bk)).
\label{gv}
\eeq
We have also assumed that the spin-quantization axis is parallel to the field. 

\vspace{10pt}

\noindent {\bf Data availability} All relevant data and codes are available from the authors upon reasonable request.

}

\acknowledgements

We thank Shiang Fang, Yahui Zhang, Yizhuang You, Erez Berg, and Bertrand Halperin for helpful discussion.
In particular, we thank Mikito Koshino for clarification on his earlier works on BLG parameters.
A. Vishwanath, J.Y. Lee and E. Khalaf were supported by a Simons Investigator Fellowship. P. Kim, X. Liu and Z. Hao acknowledge partial support from the Gordon and Betty Moore Foundation's EPiQS Initiative through Grant GBMF4543 and the DoD Vannevar Bush Faculty Fellowship N00014-18-1-2877.

\bibliography{ref.bib}

\begin{thebibliography}{10}
\expandafter\ifx\csname url\endcsname\relax
  \def\url#1{\texttt{#1}}\fi
\expandafter\ifx\csname urlprefix\endcsname\relax\def\urlprefix{URL }\fi
\providecommand{\bibinfo}[2]{#2}
\providecommand{\eprint}[2][]{\url{#2}}

\bibitem{PabloMott}
\bibinfo{author}{Cao, Y.} \emph{et~al.}
\newblock \bibinfo{title}{Correlated insulator behaviour at half-filling in
  magic-angle graphene superlattices}.
\newblock \emph{\bibinfo{journal}{Nature}} \textbf{\bibinfo{volume}{556}},
  \bibinfo{pages}{80} (\bibinfo{year}{2018}).

\bibitem{PabloSC}
\bibinfo{author}{Cao, Y.} \emph{et~al.}
\newblock \bibinfo{title}{Unconventional superconductivity in magic-angle
  graphene superlattices}.
\newblock \emph{\bibinfo{journal}{Nature}} \textbf{\bibinfo{volume}{556}},
  \bibinfo{pages}{43} (\bibinfo{year}{2018}).

\bibitem{Dean-Young}
\bibinfo{author}{Yankowitz, M.} \emph{et~al.}
\newblock \bibinfo{title}{Tuning superconductivity in twisted bilayer
  graphene}.
\newblock \emph{\bibinfo{journal}{Science}} \bibinfo{pages}{eaav1910}
  (\bibinfo{year}{2019}).

\bibitem{Efetov}
\bibinfo{author}{Lu, X.} \emph{et~al.}
\newblock \bibinfo{title}{Superconductors, orbital magnets, and correlated
  states in magic angle bilayer graphene}.
\newblock \emph{\bibinfo{journal}{arXiv preprint arXiv:1903.06513}}
  (\bibinfo{year}{2019}).

\bibitem{MacDonald}
\bibinfo{author}{Bistritzer, R.} \& \bibinfo{author}{MacDonald, A.~H.}
\newblock \bibinfo{title}{Moir{\'e} bands in twisted double-layer graphene}.
\newblock \emph{\bibinfo{journal}{Proceedings of the National Academy of
  Sciences}} \textbf{\bibinfo{volume}{108}}, \bibinfo{pages}{12233--12237}
  (\bibinfo{year}{2011}).

\bibitem{Santos}
\bibinfo{author}{Lopes~dos Santos, J. M.~B.}, \bibinfo{author}{Peres, N. M.~R.}
  \& \bibinfo{author}{Castro~Neto, A.~H.}
\newblock \bibinfo{title}{Continuum model of the twisted graphene bilayer}.
\newblock \emph{\bibinfo{journal}{Phys. Rev. B}} \textbf{\bibinfo{volume}{86}},
  \bibinfo{pages}{155449} (\bibinfo{year}{2012}).

\bibitem{Balents18}
\bibinfo{author}{Xu, C.} \& \bibinfo{author}{Balents, L.}
\newblock \bibinfo{title}{Topological superconductivity in twisted multilayer
  graphene}.
\newblock \emph{\bibinfo{journal}{Phys. Rev. Lett.}}
  \textbf{\bibinfo{volume}{121}}, \bibinfo{pages}{087001}
  (\bibinfo{year}{2018}).

\bibitem{Po2018}
\bibinfo{author}{Po, H.~C.}, \bibinfo{author}{Zou, L.},
  \bibinfo{author}{Vishwanath, A.} \& \bibinfo{author}{Senthil, T.}
\newblock \bibinfo{title}{Origin of mott insulating behavior and
  superconductivity in twisted bilayer graphene}.
\newblock \emph{\bibinfo{journal}{Phys. Rev. X}} \textbf{\bibinfo{volume}{8}},
  \bibinfo{pages}{031089} (\bibinfo{year}{2018}).

\bibitem{IsobeFu}
\bibinfo{author}{Isobe, H.}, \bibinfo{author}{Yuan, N. F.~Q.} \&
  \bibinfo{author}{Fu, L.}
\newblock \bibinfo{title}{Unconventional superconductivity and density waves in
  twisted bilayer graphene}.
\newblock \emph{\bibinfo{journal}{Phys. Rev. X}} \textbf{\bibinfo{volume}{8}},
  \bibinfo{pages}{041041} (\bibinfo{year}{2018}).

\bibitem{Thomson18}
\bibinfo{author}{Thomson, A.}, \bibinfo{author}{Chatterjee, S.},
  \bibinfo{author}{Sachdev, S.} \& \bibinfo{author}{Scheurer, M.~S.}
\newblock \bibinfo{title}{Triangular antiferromagnetism on the honeycomb
  lattice of twisted bilayer graphene}.
\newblock \emph{\bibinfo{journal}{Phys. Rev. B}} \textbf{\bibinfo{volume}{98}},
  \bibinfo{pages}{075109} (\bibinfo{year}{2018}).

\bibitem{YouAV}
\bibinfo{author}{You, Y.-Z.} \& \bibinfo{author}{Vishwanath, A.}
\newblock \bibinfo{title}{Superconductivity from valley fluctuations and
  approximate so (4) symmetry in a weak coupling theory of twisted bilayer
  graphene}.
\newblock \emph{\bibinfo{journal}{arXiv preprint arXiv:1805.06867}}
  (\bibinfo{year}{2018}).

\bibitem{Vafek}
\bibinfo{author}{Kang, J.} \& \bibinfo{author}{Vafek, O.}
\newblock \bibinfo{title}{Symmetry, maximally localized wannier states, and a
  low-energy model for twisted bilayer graphene narrow bands}.
\newblock \emph{\bibinfo{journal}{Phys. Rev. X}} \textbf{\bibinfo{volume}{8}},
  \bibinfo{pages}{031088} (\bibinfo{year}{2018}).

\bibitem{Xie2018}
\bibinfo{author}{Xie, M.} \& \bibinfo{author}{MacDonald, A.~H.}
\newblock \bibinfo{title}{On the nature of the correlated insulator states in
  twisted bilayer graphene}.
\newblock \emph{\bibinfo{journal}{arXiv preprint arXiv:1812.04213}}
  (\bibinfo{year}{2018}).

\bibitem{Nandkishore}
\bibinfo{author}{Lin, Y.-P.} \& \bibinfo{author}{Nandkishore, R.~M.}
\newblock \bibinfo{title}{A chiral twist on the high-$ t\_c $ phase diagram in
  moir$\backslash$'e heterostructures}.
\newblock \emph{\bibinfo{journal}{arXiv preprint arXiv:1901.00500}}
  (\bibinfo{year}{2019}).

\bibitem{Kivelson}
\bibinfo{author}{Dodaro, J.~F.}, \bibinfo{author}{Kivelson, S.~A.},
  \bibinfo{author}{Schattner, Y.}, \bibinfo{author}{Sun, X.~Q.} \&
  \bibinfo{author}{Wang, C.}
\newblock \bibinfo{title}{Phases of a phenomenological model of twisted bilayer
  graphene}.
\newblock \emph{\bibinfo{journal}{Phys. Rev. B}} \textbf{\bibinfo{volume}{98}},
  \bibinfo{pages}{075154} (\bibinfo{year}{2018}).

\bibitem{PhilipPhillips}
\bibinfo{author}{Padhi, B.}, \bibinfo{author}{Setty, C.} \&
  \bibinfo{author}{Phillips, P.~W.}
\newblock \bibinfo{title}{Doped twisted bilayer graphene near magic angles:
  Proximity to wigner crystallization, not mott insulation}.
\newblock \emph{\bibinfo{journal}{Nano letters}} \textbf{\bibinfo{volume}{18}},
  \bibinfo{pages}{6175--6180} (\bibinfo{year}{2018}).

\bibitem{phononMacDonald}
\bibinfo{author}{Wu, F.}, \bibinfo{author}{MacDonald, A.~H.} \&
  \bibinfo{author}{Martin, I.}
\newblock \bibinfo{title}{Theory of phonon-mediated superconductivity in
  twisted bilayer graphene}.
\newblock \emph{\bibinfo{journal}{Phys. Rev. Lett.}}
  \textbf{\bibinfo{volume}{121}}, \bibinfo{pages}{257001}
  (\bibinfo{year}{2018}).

\bibitem{phononLianBernevig}
\bibinfo{author}{Lian, B.}, \bibinfo{author}{Wang, Z.} \&
  \bibinfo{author}{Bernevig, B.~A.}
\newblock \bibinfo{title}{Twisted bilayer graphene: A phonon driven
  superconductor}.
\newblock \emph{\bibinfo{journal}{arXiv preprint arXiv:1807.04382}}
  (\bibinfo{year}{2018}).

\bibitem{Zou2018}
\bibinfo{author}{Zou, L.}, \bibinfo{author}{Po, H.~C.},
  \bibinfo{author}{Vishwanath, A.} \& \bibinfo{author}{Senthil, T.}
\newblock \bibinfo{title}{Band structure of twisted bilayer graphene: Emergent
  symmetries, commensurate approximants, and wannier obstructions}.
\newblock \emph{\bibinfo{journal}{Phys. Rev. B}} \textbf{\bibinfo{volume}{98}},
  \bibinfo{pages}{085435} (\bibinfo{year}{2018}).

\bibitem{TDBGexp2019}
\bibinfo{author}{Liu, X.} \emph{et~al.}
\newblock \bibinfo{title}{Spin-polarized correlated insulator and
  superconductorin twisted double bilayer graphene}
  \bibinfo{pages}{arXiv:1903.08130} (\bibinfo{year}{2019}).

\bibitem{Koshino2017}
\bibinfo{author}{Nam, N. N.~T.} \& \bibinfo{author}{Koshino, M.}
\newblock \bibinfo{title}{Lattice relaxation and energy band modulation in
  twisted bilayer graphene}.
\newblock \emph{\bibinfo{journal}{Phys. Rev. B}} \textbf{\bibinfo{volume}{96}},
  \bibinfo{pages}{075311} (\bibinfo{year}{2017}).

\bibitem{Koshino2018}
\bibinfo{author}{Koshino, M.} \emph{et~al.}
\newblock \bibinfo{title}{Maximally localized wannier orbitals and the extended
  hubbard model for twisted bilayer graphene}.
\newblock \emph{\bibinfo{journal}{Phys. Rev. X}} \textbf{\bibinfo{volume}{8}},
  \bibinfo{pages}{031087} (\bibinfo{year}{2018}).

\bibitem{Goerbig}
\bibinfo{author}{Goerbig, M.~O.}
\newblock \bibinfo{title}{Electronic properties of graphene in a strong
  magnetic field}.
\newblock \emph{\bibinfo{journal}{Rev. Mod. Phys.}}
  \textbf{\bibinfo{volume}{83}}, \bibinfo{pages}{1193--1243}
  (\bibinfo{year}{2011}).

\bibitem{Zhang2018}
\bibinfo{author}{Zhang, Y.-H.}, \bibinfo{author}{Mao, D.},
  \bibinfo{author}{Cao, Y.}, \bibinfo{author}{Jarillo-Herrero, P.} \&
  \bibinfo{author}{Senthil, T.}
\newblock \bibinfo{title}{Nearly flat chern bands in moir\'e superlattices}.
\newblock \emph{\bibinfo{journal}{Phys. Rev. B}} \textbf{\bibinfo{volume}{99}},
  \bibinfo{pages}{075127} (\bibinfo{year}{2019}).

\bibitem{Crommie2010}
\bibinfo{author}{Levy, N.} \emph{et~al.}
\newblock \bibinfo{title}{Strain-induced pseudo{\textendash}magnetic fields
  greater than 300 tesla in graphene nanobubbles}.
\newblock \emph{\bibinfo{journal}{Science}} \textbf{\bibinfo{volume}{329}},
  \bibinfo{pages}{544--547} (\bibinfo{year}{2010}).

\bibitem{Ghaemi2012}
\bibinfo{author}{Ghaemi, P.}, \bibinfo{author}{Cayssol, J.},
  \bibinfo{author}{Sheng, D.~N.} \& \bibinfo{author}{Vishwanath, A.}
\newblock \bibinfo{title}{Fractional topological phases and broken
  time-reversal symmetry in strained graphene}.
\newblock \emph{\bibinfo{journal}{Phys. Rev. Lett.}}
  \textbf{\bibinfo{volume}{108}}, \bibinfo{pages}{266801}
  (\bibinfo{year}{2012}).

\bibitem{Sharpe2019}
\bibinfo{author}{{Sharpe}, A.~L.} \emph{et~al.}
\newblock \bibinfo{title}{{Emergent ferromagnetism near three-quarters filling
  in twisted bilayer graphene}}.
\newblock \emph{\bibinfo{journal}{arXiv e-prints}}
  \bibinfo{pages}{arXiv:1901.03520} (\bibinfo{year}{2019}).

\bibitem{Zhang2019}
\bibinfo{author}{{Zhang}, Y.-H.}, \bibinfo{author}{{Mao}, D.} \&
  \bibinfo{author}{{Senthil}, T.}
\newblock \bibinfo{title}{{Twisted Bilayer Graphene Aligned with Hexagonal
  Boron Nitride: Anomalous Hall Effect and a Lattice Model}}.
\newblock \emph{\bibinfo{journal}{arXiv e-prints}}
  \bibinfo{pages}{arXiv:1901.08209} (\bibinfo{year}{2019}).

\bibitem{Bultinck19}
\bibinfo{author}{Bultinck, N.}, \bibinfo{author}{Chatterjee, S.} \&
  \bibinfo{author}{Zaletel, M.~P.}
\newblock \bibinfo{title}{Anomalous hall ferromagnetism in twisted bilayer
  graphene}.
\newblock \emph{\bibinfo{journal}{arXiv preprint arXiv:1901.08110}}
  (\bibinfo{year}{2019}).

\bibitem{MacDonald2011}
\bibinfo{author}{Bistritzer, R.} \& \bibinfo{author}{MacDonald, A.~H.}
\newblock \bibinfo{title}{Moir{\'e} bands in twisted double-layer graphene}.
\newblock \emph{\bibinfo{journal}{Proceedings of the National Academy of
  Sciences}} \textbf{\bibinfo{volume}{108}}, \bibinfo{pages}{12233--12237}
  (\bibinfo{year}{2011}).

\bibitem{Tarnopolsky}
\bibinfo{author}{Tarnopolsky, G.}, \bibinfo{author}{Kruchkov, A.~J.} \&
  \bibinfo{author}{Vishwanath, A.}
\newblock \bibinfo{title}{Origin of magic angles in twisted bilayer graphene}.
\newblock \emph{\bibinfo{journal}{Phys. Rev. Lett.}}
  \textbf{\bibinfo{volume}{122}}, \bibinfo{pages}{106405}
  (\bibinfo{year}{2019}).

\bibitem{Khalaf2019}
\bibinfo{author}{{Khalaf}, E.}, \bibinfo{author}{{Kruchkov}, A.~J.},
  \bibinfo{author}{{Tarnopolsky}, G.} \& \bibinfo{author}{{Vishwanath}, A.}
\newblock \bibinfo{title}{{Magic Angle Hierarchy in Twisted Graphene
  Multilayers}}.
\newblock \emph{\bibinfo{journal}{arXiv e-prints}}
  \bibinfo{pages}{arXiv:1901.10485} (\bibinfo{year}{2019}).

\bibitem{ChoiSTM2019}
\bibinfo{author}{{Choi}, Y.} \emph{et~al.}
\newblock \bibinfo{title}{{Imaging Electronic Correlations in Twisted Bilayer
  Graphene near the Magic Angle}}.
\newblock \emph{\bibinfo{journal}{arXiv e-prints}}
  \bibinfo{pages}{arXiv:1901.02997} (\bibinfo{year}{2019}).

\bibitem{BLG_valley1}
\bibinfo{author}{Martin, I.}, \bibinfo{author}{Blanter, Y.~M.} \&
  \bibinfo{author}{Morpurgo, A.~F.}
\newblock \bibinfo{title}{Topological confinement in bilayer graphene}.
\newblock \emph{\bibinfo{journal}{Phys. Rev. Lett.}}
  \textbf{\bibinfo{volume}{100}}, \bibinfo{pages}{036804}
  (\bibinfo{year}{2008}).

\bibitem{BLG_valley2}
\bibinfo{author}{Vaezi, A.}, \bibinfo{author}{Liang, Y.},
  \bibinfo{author}{Ngai, D.~H.}, \bibinfo{author}{Yang, L.} \&
  \bibinfo{author}{Kim, E.-A.}
\newblock \bibinfo{title}{Topological edge states at a tilt boundary in gated
  multilayer graphene}.
\newblock \emph{\bibinfo{journal}{Phys. Rev. X}} \textbf{\bibinfo{volume}{3}},
  \bibinfo{pages}{021018} (\bibinfo{year}{2013}).

\bibitem{BLG_valley3}
\bibinfo{author}{Zhang, F.}, \bibinfo{author}{MacDonald, A.~H.} \&
  \bibinfo{author}{Mele, E.~J.}
\newblock \bibinfo{title}{Valley chern numbers and boundary modes in gapped
  bilayer graphene}.
\newblock \emph{\bibinfo{journal}{Proceedings of the National Academy of
  Sciences}} \textbf{\bibinfo{volume}{110}}, \bibinfo{pages}{10546--10551}
  (\bibinfo{year}{2013}).
\newblock arXiv:\eprint{https://www.pnas.org/content/110/26/10546.full.pdf}.

\bibitem{BLG_valley_exp}
\bibinfo{author}{Ju, L.} \emph{et~al.}
\newblock \bibinfo{title}{Topological valley transport at bilayer graphene
  domain walls}.
\newblock \emph{\bibinfo{journal}{Nature}} \textbf{\bibinfo{volume}{520}},
  \bibinfo{pages}{650 EP --} (\bibinfo{year}{2015}).

\bibitem{Koshino2011}
\bibinfo{author}{Koshino, M.}
\newblock \bibinfo{title}{Chiral orbital current and anomalous magnetic moment
  in gapped graphene}.
\newblock \emph{\bibinfo{journal}{Phys. Rev. B}} \textbf{\bibinfo{volume}{84}},
  \bibinfo{pages}{125427} (\bibinfo{year}{2011}).

\bibitem{McEuen2017}
\bibinfo{author}{Ju, L.} \emph{et~al.}
\newblock \bibinfo{title}{Tunable excitons in bilayer graphene}.
\newblock \emph{\bibinfo{journal}{Science}} \textbf{\bibinfo{volume}{358}},
  \bibinfo{pages}{907--910} (\bibinfo{year}{2017}).

\bibitem{Vanderbilt1997}
\bibinfo{author}{Marzari, N.} \& \bibinfo{author}{Vanderbilt, D.}
\newblock \bibinfo{title}{Maximally localized generalized wannier functions for
  composite energy bands}.
\newblock \emph{\bibinfo{journal}{Phys. Rev. B}} \textbf{\bibinfo{volume}{56}},
  \bibinfo{pages}{12847--12865} (\bibinfo{year}{1997}).

\bibitem{Young_FCI_2018}
\bibinfo{author}{Spanton, E.~M.} \emph{et~al.}
\newblock \bibinfo{title}{Observation of fractional chern insulators in a van
  der waals heterostructure}.
\newblock \emph{\bibinfo{journal}{Science}} \textbf{\bibinfo{volume}{360}},
  \bibinfo{pages}{62--66} (\bibinfo{year}{2018}).
\newblock
  arXiv:\eprint{https://science.sciencemag.org/content/360/6384/62.full.pdf}.

\bibitem{Screening_2}
\bibinfo{author}{Papi\ifmmode~\acute{c}\else \'{c}\fi{}, Z.} \&
  \bibinfo{author}{Abanin, D.~A.}
\newblock \bibinfo{title}{Topological phases in the zeroth landau level of
  bilayer graphene}.
\newblock \emph{\bibinfo{journal}{Phys. Rev. Lett.}}
  \textbf{\bibinfo{volume}{112}}, \bibinfo{pages}{046602}
  (\bibinfo{year}{2014}).

\bibitem{SC_TMD_theory1}
\bibinfo{author}{Ge, Y.} \& \bibinfo{author}{Liu, A.~Y.}
\newblock \bibinfo{title}{Phonon-mediated superconductivity in electron-doped
  single-layer mos${}_{2}$: A first-principles prediction}.
\newblock \emph{\bibinfo{journal}{Phys. Rev. B}} \textbf{\bibinfo{volume}{87}},
  \bibinfo{pages}{241408} (\bibinfo{year}{2013}).

\bibitem{SC_TMD_theory2}
\bibinfo{author}{Rold\'an, R.}, \bibinfo{author}{Cappelluti, E.} \&
  \bibinfo{author}{Guinea, F.}
\newblock \bibinfo{title}{Interactions and superconductivity in heavily doped
  mos${}_{2}$}.
\newblock \emph{\bibinfo{journal}{Phys. Rev. B}} \textbf{\bibinfo{volume}{88}},
  \bibinfo{pages}{054515} (\bibinfo{year}{2013}).

\bibitem{SC_TMD_theory3}
\bibinfo{author}{Yuan, N. F.~Q.}, \bibinfo{author}{Mak, K.~F.} \&
  \bibinfo{author}{Law, K.~T.}
\newblock \bibinfo{title}{Possible topological superconducting phases of
  ${\mathrm{mos}}_{2}$}.
\newblock \emph{\bibinfo{journal}{Phys. Rev. Lett.}}
  \textbf{\bibinfo{volume}{113}}, \bibinfo{pages}{097001}
  (\bibinfo{year}{2014}).

\bibitem{SC_TMD_exp}
\bibinfo{author}{Saito, Y.} \emph{et~al.}
\newblock \bibinfo{title}{Superconductivity protected by spin--valley locking
  in ion-gated mos2}.
\newblock \emph{\bibinfo{journal}{Nature Physics}}
  \textbf{\bibinfo{volume}{12}}, \bibinfo{pages}{144 EP --}
  (\bibinfo{year}{2015}).

\bibitem{saxena_superconductivity_2000}
\bibinfo{author}{Saxena, S.~S.} \emph{et~al.}
\newblock \bibinfo{title}{Superconductivity on the border of itinerant-electron
  ferromagnetism in {UGe}$_{\textrm{2}}$}.
\newblock \emph{\bibinfo{journal}{Nature}} \textbf{\bibinfo{volume}{406}},
  \bibinfo{pages}{587--592} (\bibinfo{year}{2000}).

\bibitem{aoki_coexistence_2001}
\bibinfo{author}{Aoki, D.} \emph{et~al.}
\newblock \bibinfo{title}{Coexistence of superconductivity and ferromagnetism
  in {URhGe}}.
\newblock \emph{\bibinfo{journal}{Nature}} \textbf{\bibinfo{volume}{413}},
  \bibinfo{pages}{613--616} (\bibinfo{year}{2001}).

\bibitem{huy_superconductivity_2007}
\bibinfo{author}{Huy, N.~T.} \emph{et~al.}
\newblock \bibinfo{title}{Superconductivity on the {Border} of {Weak}
  {Itinerant} {Ferromagnetism} in {UCoGe}}.
\newblock \emph{\bibinfo{journal}{Physical Review Letters}}
  \textbf{\bibinfo{volume}{99}} (\bibinfo{year}{2007}).

\bibitem{TBG_SC_triplet2018}
\bibinfo{author}{Liu, C.-C.}, \bibinfo{author}{Zhang, L.-D.},
  \bibinfo{author}{Chen, W.-Q.} \& \bibinfo{author}{Yang, F.}
\newblock \bibinfo{title}{Chiral spin density wave and $d+id$ superconductivity
  in the magic-angle-twisted bilayer graphene}.
\newblock \emph{\bibinfo{journal}{Phys. Rev. Lett.}}
  \textbf{\bibinfo{volume}{121}}, \bibinfo{pages}{217001}
  (\bibinfo{year}{2018}).

\bibitem{Ambegaokar73}
\bibinfo{author}{Ambegaokar, V.} \& \bibinfo{author}{Mermin, N.~D.}
\newblock \bibinfo{title}{Thermal anomalies of ${\mathrm{he}}^{3}$: Pairing in
  a magnetic field}.
\newblock \emph{\bibinfo{journal}{Phys. Rev. Lett.}}
  \textbf{\bibinfo{volume}{30}}, \bibinfo{pages}{81--84}
  (\bibinfo{year}{1973}).

\bibitem{Herz76}
\bibinfo{author}{Hertz, J.~A.}
\newblock \bibinfo{title}{Quantum critical phenomena}.
\newblock \emph{\bibinfo{journal}{Phys. Rev. B}} \textbf{\bibinfo{volume}{14}},
  \bibinfo{pages}{1165--1184} (\bibinfo{year}{1976}).

\bibitem{Millis93}
\bibinfo{author}{Millis, A.~J.}
\newblock \bibinfo{title}{Effect of a nonzero temperature on quantum critical
  points in itinerant fermion systems}.
\newblock \emph{\bibinfo{journal}{Phys. Rev. B}} \textbf{\bibinfo{volume}{48}},
  \bibinfo{pages}{7183--7196} (\bibinfo{year}{1993}).

\bibitem{Hattori11}
\bibinfo{author}{Hattori, T.} \emph{et~al.}
\newblock \bibinfo{title}{Superconductivity induced by longitudinal
  ferromagnetic fluctuations in ucoge}.
\newblock \emph{\bibinfo{journal}{Phys. Rev. Lett.}}
  \textbf{\bibinfo{volume}{108}}, \bibinfo{pages}{066403}
  (\bibinfo{year}{2012}).

\bibitem{PabloTDBG}
\bibinfo{author}{{Cao}, Y.} \emph{et~al.}
\newblock \bibinfo{title}{{Electric Field Tunable Correlated States and
  Magnetic Phase Transitions in Twisted Bilayer-Bilayer Graphene}}.
\newblock \emph{\bibinfo{journal}{arXiv e-prints}}
  \bibinfo{pages}{arXiv:1903.08596} (\bibinfo{year}{2019}).

\bibitem{IOP_TDBG}
\bibinfo{author}{{Shen}, C.} \emph{et~al.}
\newblock \bibinfo{title}{{Observation of superconductivity with Tc onset at
  12K in electrically tunable twisted double bilayer graphene}}.
\newblock \emph{\bibinfo{journal}{arXiv e-prints}}
  \bibinfo{pages}{arXiv:1903.06952} (\bibinfo{year}{2019}).

\bibitem{Jeil2014}
\bibinfo{author}{Jung, J.} \& \bibinfo{author}{MacDonald, A.~H.}
\newblock \bibinfo{title}{Accurate tight-binding models for the
  $\ensuremath{\pi}$ bands of bilayer graphene}.
\newblock \emph{\bibinfo{journal}{Phys. Rev. B}} \textbf{\bibinfo{volume}{89}},
  \bibinfo{pages}{035405} (\bibinfo{year}{2014}).

\bibitem{Bernevig2012}
\bibinfo{author}{Fang, C.}, \bibinfo{author}{Gilbert, M.~J.} \&
  \bibinfo{author}{Bernevig, B.~A.}
\newblock \bibinfo{title}{Bulk topological invariants in noninteracting point
  group symmetric insulators}.
\newblock \emph{\bibinfo{journal}{Phys. Rev. B}} \textbf{\bibinfo{volume}{86}},
  \bibinfo{pages}{115112} (\bibinfo{year}{2012}).

\bibitem{Turner2012}
\bibinfo{author}{{Turner}, A.~M.}, \bibinfo{author}{{Zhang}, Y.},
  \bibinfo{author}{{Mong}, R. S.~K.} \& \bibinfo{author}{{Vishwanath}, A.}
\newblock \bibinfo{title}{{Quantized response and topology of magnetic
  insulators with inversion symmetry}}.
\newblock \emph{\bibinfo{journal}{Physical Review B}}
  \textbf{\bibinfo{volume}{85}}, \bibinfo{pages}{165120}
  (\bibinfo{year}{2012}).

\bibitem{Watanabe2017}
\bibinfo{author}{Matsugatani, A.}, \bibinfo{author}{Ishiguro, Y.},
  \bibinfo{author}{Shiozaki, K.} \& \bibinfo{author}{Watanabe, H.}
\newblock \bibinfo{title}{Universal relation among the many-body chern number,
  rotation symmetry, and filling}.
\newblock \emph{\bibinfo{journal}{Phys. Rev. Lett.}}
  \textbf{\bibinfo{volume}{120}}, \bibinfo{pages}{096601}
  (\bibinfo{year}{2018}).

\bibitem{SlaterKoster1954}
\bibinfo{author}{Slater, J.~C.} \& \bibinfo{author}{Koster, G.~F.}
\newblock \bibinfo{title}{Simplified lcao method for the periodic potential
  problem}.
\newblock \emph{\bibinfo{journal}{Phys. Rev.}} \textbf{\bibinfo{volume}{94}},
  \bibinfo{pages}{1498--1524} (\bibinfo{year}{1954}).

\bibitem{McCann2013}
\bibinfo{author}{McCann, E.} \& \bibinfo{author}{Koshino, M.}
\newblock \bibinfo{title}{The electronic properties of bilayer graphene}.
\newblock \emph{\bibinfo{journal}{Reports on Progress in Physics}}
  \textbf{\bibinfo{volume}{76}}, \bibinfo{pages}{056503}
  (\bibinfo{year}{2013}).

\bibitem{Charlier1991}
\bibinfo{author}{Charlier, J.-C.}, \bibinfo{author}{Gonze, X.} \&
  \bibinfo{author}{Michenaud, J.-P.}
\newblock \bibinfo{title}{First-principles study of the electronic properties
  of graphite}.
\newblock \emph{\bibinfo{journal}{Phys. Rev. B}} \textbf{\bibinfo{volume}{43}},
  \bibinfo{pages}{4579--4589} (\bibinfo{year}{1991}).

\bibitem{Moon2013}
\bibinfo{author}{Moon, P.} \& \bibinfo{author}{Koshino, M.}
\newblock \bibinfo{title}{Optical absorption in twisted bilayer graphene}.
\newblock \emph{\bibinfo{journal}{Phys. Rev. B}} \textbf{\bibinfo{volume}{87}},
  \bibinfo{pages}{205404} (\bibinfo{year}{2013}).

\bibitem{Kohmoto1985}
\bibinfo{author}{Kohmoto, M.}
\newblock \bibinfo{title}{Topological invariant and the quantization of the
  hall conductance}.
\newblock \emph{\bibinfo{journal}{Annals of Physics}}
  \textbf{\bibinfo{volume}{160}}, \bibinfo{pages}{343 -- 354}
  (\bibinfo{year}{1985}).

\bibitem{Carr2019}
\bibinfo{author}{{Carr}, S.}, \bibinfo{author}{{Fang}, S.},
  \bibinfo{author}{{Zhu}, Z.} \& \bibinfo{author}{{Kaxiras}, E.}
\newblock \bibinfo{title}{{Minimal model for low-energy electronic states of
  twisted bilayer graphene}}.
\newblock \emph{\bibinfo{journal}{arXiv e-prints}}
  \bibinfo{pages}{arXiv:1901.03420} (\bibinfo{year}{2019}).

\bibitem{usov1988theory}
\bibinfo{author}{Usov, N.}
\newblock \bibinfo{title}{Theory of the quantum hall effect in a
  two-dimensional periodic potential}.
\newblock \emph{\bibinfo{journal}{JETP}} \textbf{\bibinfo{volume}{67}},
  \bibinfo{pages}{2565} (\bibinfo{year}{1988}).

\end{thebibliography}

\setcounter{equation}{0}
\setcounter{section}{0}
\setcounter{figure}{0}
\renewcommand{\theequation}{S\arabic{equation}}
\renewcommand{\thefigure}{S\arabic{figure}}

\begin{widetext}

\begin{center}

{\large \bf \noindent
Supplementary Material}

\vskip4mm

{\bf \noindent
Theory of correlated insulating behaviour and spin-triplet superconductivity in twisted double bilayer graphene}

\vskip3mm

{\noindent
Jong Yeon Lee\,$^*$, Eslam Khalaf\,$^*$, Shang Liu, Xiaomeng Liu, Zeyu Hao, Philip Kim, Ashvin Vishwanath}

\vskip3mm

\end{center}

In this supplementary material, we present detailed numerical and analytical methods employed to obtain the results discussed in the main article. First, we elaborate on the calculation of the TDBG Moire band structure. Second, we discuss how intervalley Hund's coupling arises from the projection of the Coulomb interaction on the first conduction band. Third, we present the self-consistent Hartree-Fock method, which was carried out accurately compared to Ref.~\onlinecite{Zhang2018,Po2018}. Fourth, we study the competition between IVC and VP in two bands with opposite Chern number using a perturbative Hartree-Fock calculation. Finally, we study the dependence of $T_c$ on magnetic field.  
\vskip5mm

\tableofcontents

\section{Details of Twisted Double Bilayer Graphene Continuum Model}\label{app:convention}

Before getting into the detailed calculation, here we clarify some subtleties in the Moir\'e continuum approach and tight-binding parameters. As we will see, this is crucial because it can affect the physical band structures of a twisted double bilayer graphene. First, let us fix the lattice convention. Let $a=1.42 \textrm{ \AA}$ be the distance between carbon atoms. The original hexagonal lattice is defined by 
\beq
a_1 = a (\sqrt{3},0), \quad a_2 = a (-\frac{\sqrt{3}}{2},\frac{3}{2} ) \quad \Leftrightarrow \quad G_1 = \frac{4\pi}{3a}  ( \frac{\sqrt{3}}{2}, \frac{1}{2} ), \quad G_2 = \frac{4\pi}{3a} (0, 1), \quad G_3 = G_2 - G_1
\eeq
Here, $\bK_\pm$-point is given by $\frac{4\pi}{3\sqrt{3} a}(\pm 1,0)$.  Once we twist two layers, Moir\'e structure is formed with a spatial modulation given by linear combinations of $\{ {\cal R}_{\theta/2} G_1, {\cal R}_{\theta/2} G_2, {\cal R}_{-\theta/2}G_1, {\cal R}_{-\theta/2} G_2\} $, where ${\cal R}_\theta$ rotates a vector by angle $\theta$ counterclockwise. Periodicity of the given system is then governed by the smallest reciprocal lattice vector that can be obtained by linear combination. These are $({\cal R}_{\theta/2} - {\cal R}_{-\theta/2})G_1$ and $({\cal R}_{\theta/2} - {\cal R}_{-\theta/2})G_2$, which correspond to Moir\'e reciprocal lattice vectors. Therefore, we obtain $G_{1,2}^M = \frac{8\pi \sin \theta/2}{{3} a} (-1/2, \pm \sqrt{3}/2)$. For convenience, use the following Moir\'e reciprocal lattice vectors from now on:
\begin{equation}
    G_{1}^M = \frac{8\pi \sin \frac{\theta}{2}  }{{3} a} \qty(\frac{1}{2}, \frac{\sqrt{3}}{2}), \quad G_{2}^M = \frac{8\pi \sin \frac{\theta}{2}  }{{3} a} \qty(-\frac{1}{2}, \frac{\sqrt{3}}{2}), \quad 
    a_{1}^M = \frac{\sqrt{3}a}{2 \sin \frac{\theta}{2}  } \qty(\frac{\sqrt{3}}{2}, \frac{1}{2}), \quad a_{2}^M =  \frac{\sqrt{3}a}{2 \sin \frac{\theta}{2}  } \qty(-s\frac{\sqrt{3}}{2}, \frac{1}{2})
\end{equation}\\

Before getting into detail, let us fix the Fourier transform convention by $c^\dagger_\bk = \sum_\bR e^{i \bk \cdot \bR} c^\dagger_\bR$. Here, $c^\dagger_\bR$ creates a Wannier orbital $W(\br-\bR)$ centered at $\bR$.
This is consistent with the other convention used throughout the paper, where $\ket{\psi_\bk} = \sum_R e^{i\bk \cdot \bR} \ket{\bR}$. 
Under this choice, whenever there is a hopping from $\bR$ to $\bR'$ in real space, one obtains the term proportional to $e^{-i \bk \cdot (\bR'-\bR)}$ in the Bloch Hamiltonian. 
Here, the tight-binding model for bilayer graphene can be fully characterized by the parameters $(\gamma_0, \gamma_1, \gamma_3, \gamma_4, \Delta)$, where the hopping term for nearest neighbor $\gamma_0$ is intentionally taken to have an additional minus sign from the hopping integral so that all $\gamma_i$'s are positive. The sign difference between vertical ($V_{pp\pi}$) and horizontal hopping ($V_{pp\sigma}$) overlap integral is originated from the phase structure of $2 p_z$ orbital.  (See how Slater-Koster parameters \cite{SlaterKoster1954} are calculated) In Ref.~\cite{McCann2013}, the sign of the trigonal warping $\gamma_3$ is taken to be negative, which is an inaccurate choice of the parameter because different sign convention would flip the shape of trigonal warping. Following the DFT result in Ref.~\cite{Charlier1991}, positive sign in front of $\gamma_3$ should be a proper choice for realistic materials.

\begin{figure}[t]
\begin{center}
\includegraphics[width=0.85\columnwidth]{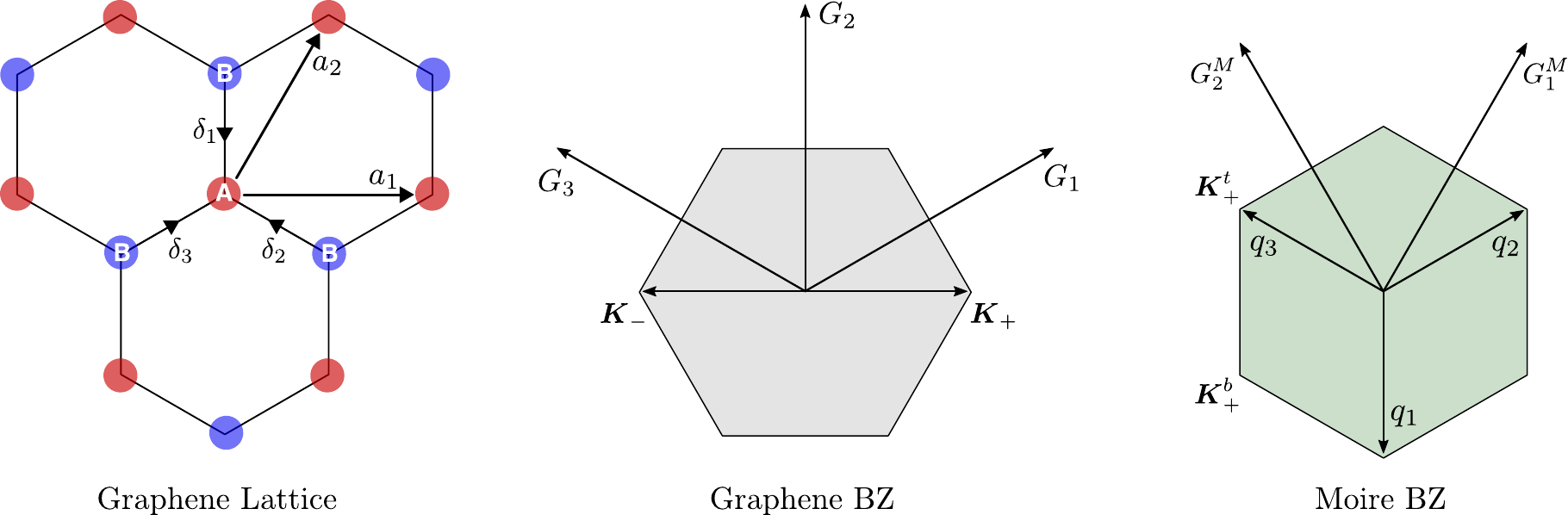}
\caption{ Real space lattice and Brillouin zone for a single graphene layer, and Brillouin zone origniated from $\boldsymbol{K}_+$ valley for a twisted (double) bilayer graphene. The figure illustrates all vectors labeled for the discussion in the manuscript. }
\label{fig:sup}
\end{center}
\end{figure}

Under this convention, where the phase structure for $2p_z$ orbitals at every carbon site is taken to be equivalent, one can derive the Moir\'e hopping term as the following. Once we have a Moir\'e structure, between momentum points of top and bottom layers, there exists Moir\'e-hopping term whose momentum transfer is given by linear combinations of $G^M_1$ and $G^M_2$. As we are interested in a Moire band structure near charge neutral point, we only consider electron momenta near Dirac points for top and bottom layers, ${\cal R}_{\pm \theta/2} \bK_\pm$-points. For example, with respect to ${\cal R}_{\pm \theta/2} \bK_+$, the momentum transfer condition can be written as
\begin{equation}
    k_\textrm{top} -     k_\textrm{bot} \equiv  0 \mod G_{1,2}^M \qquad \Rightarrow \qquad (k_\textrm{top} - {\cal R}_{\theta/2}\bK_+) - (k_\textrm{bot} - {\cal R}_{-\theta/2} \bK_+) \equiv q_1 \mod G_{1,2}^M,
\end{equation}
where $q_1 = ({\cal R}_{-\theta/2} - {\cal R}_{\theta/2} ) \bK_+ = K_M(0,-1)$ with  ${K_M} = \frac{8\pi \sin \frac{\theta}{2}}{3\sqrt{3} a}$. 
Let us denote $\bK_+^t = \qty({\cal R}_{\theta/2} \bK_+)$ and $\bK_+^b = \qty({\cal R}_{-\theta/2} \bK_+)$. The Moir\'e state with a momentum $p$ is given by a superposition of Bloch states of top and bottom layers with (absolute) momenta $\{p + n_1 G_1^M + n_2 G_2^M|n_1,n_2\in \mathbb{Z} \}$. Since we want to solve the equation in terms of Dirac Hamiltonians with respects to $\bK_+^{b,t}$-points, we do the following precedure. Define $\tilde{p} = p - \bK_+^t$. Then, the Moir\'e state with momenta $p$ is composed of the Bloch states with the following momenta defined with respect to $\bK_+^t$ and $\bK_+^b$:
\begin{eqnarray}
    \textrm{Top Layer} : \{ \tilde{p}, \tilde{p}-G_1^M, \tilde{p}-G_2^M, \tilde{p} - G_1^M + G_2^M, \dots \} \quad \textrm{with respect to $\bK^t_+$} \nonumber \\
    \textrm{Bottom Layer} : \{ \tilde{p} - q_1, \tilde{p} - q_2, \tilde{p} - q_3, \tilde{p} - q_2 + G_2^M, \dots \} \quad \textrm{with respect to $\bK^b_+$} 
\end{eqnarray}
where $q_2 = (G_1^M + q_1) = K_M(\sqrt{3}/2,1/2)$ and $q_3 = (G_2^M + q_1) = K_M(-\sqrt{3}/2,1/2)$. Therefore, the state with a momentum $p = \tilde{p} + \bK_+^t$ can be solved by considering coupling among  bloch states from $H({\cal R}_{-\theta/2}[\tilde{p}-K_\textrm{lat}^t])$ and $H({\cal R}_{\theta/2}[\tilde{p}-K_\textrm{lat}^b])$. Here, $K_\textrm{lat}^{t,b}$ is a vector denoting the location of each lattice point in $\{ n_1 G_1^M + n_2 G_2^M: n_{1,2} \in \mathbb{Z}\}$ for the top layer and $\{ q_1 + n_1 G_1^M + n_2 G_2^M: n_{1,2} \in \mathbb{Z}\}$ for the bottom layer in the $k$-space.\\

Let top and bottom layer have relative shift $\delta$. For convenience, take a frame where the bottom layer is fixed and the top layer is rotated by $\theta$. Then, Each original Bloch state is represented as 
\begin{equation}
    \ket{\psi^t_{p,\beta}} = \frac{1}{\sqrt{N}} \sum_R e^{i p \cdot (R' + \tau'_\beta) } \ket{R'+\tau'_\beta} , \quad  \ket{\psi^b_{k,\alpha}} = \frac{1}{\sqrt{N}} \sum_{R'} e^{i k \cdot (R + \tau_\alpha) } \ket{R+\tau_\alpha}
\end{equation}
where $\alpha, \beta$ are sublattice indices and $\tau_\alpha$ is associated displacement. Here, primed coordinate $R' = {\cal R}_\theta (R + \delta)$ and $\tau' = {\cal R}_\theta \tau$ are for the top layer, meaning that it is rotated by $\theta$ along counter-clockwise direction. Here, $R = n_1 a_1 + n_2 a_2 $. The initial displacement $\delta$ between layers is not important in the end, as we will see. Let $\tau_0 = a(0,1)$. 
In the mono-mono case $\delta = 0$ for $A-A$ stacking and $\delta = \tau_0$ for $A-B$ stacking. $A-B$ stacking means that $A$-site of the top layer is placed at the location of $B$-site of the bottom layer.

By definition, $A$-site is original site spanning the lattice, and $B$-site is displaced by $\tau_0$ with respect to the $A$-site. 
In the bilayer(AB)-bilayer(AB)  case, $\delta = -\tau_0$ because between layers it is $BA$ stacking. For the following calculation, we take $\delta = -\tau_0$ and $d=0$.  Now, hopping matrix element from top second layer to bottom first layer $H_{b,t}$ can be evaluated by 
\begin{eqnarray}
    T^{\alpha \beta}_{k p'} &=& \bra{\psi^b_{k\alpha}} H_T \ket{\psi^t_{p'\beta}} = \frac{1}{N} \sum_{R,R'} e^{-ik(R+\tau_\alpha) + ip (R'+\tau_\beta')}   \bra{R+ \tau_\alpha}{H_T}\ket{R'+\tau_\beta'} \nonumber \\
    &=& \bra{\psi^b_{k\alpha}} H_T \ket{\psi^t_{p'\beta}} = \frac{1}{N} \sum_{R,R'} e^{-ik(R+\tau_\alpha) + ip (R'+\tau_\beta')}  \cdot  t(R+\tau_\alpha - R' - \tau_\beta) \nonumber \\
    &=& \frac{1}{N} \sum_{R,R'} e^{-ik(R+\tau_\alpha) + ip (R'+\tau_\beta')}  \frac{1}{\Omega}  \int d^2 q \, t_q e^{iq \cdot (R+\tau_\alpha - R' - \tau_\beta')} \nonumber \\
    &=& \frac{1}{N\Omega } \int d^2 q \sum_{R,R'} t_q e^{-i(k-q) \cdot (R+\tau_\alpha) + i(p-q) (R'+\tau_\beta')} \nonumber \\
    &=& \sum_{q \in \textrm{BZ}}  \,t_q\,  e^{-iG_1 \cdot \tau_\alpha + iG'_2 (\tau_\beta' + \delta')} \textrm{ non-zero only when }{k-q} = G_1 \in {\cal G} \textrm{ and } p-q = G'_2 \in {\cal G}' \nonumber \\
    &=& \sum_{g_1, g_2}  \,t_{k-g_1}\,  e^{-ig_1 \cdot \tau_\alpha + ig_2 \cdot (\tau_\beta + \delta)} \cdot \delta_{p+g_1,k+g'_2} \nonumber \\
    &=& \sum_{g_1, g_2}  \,t_{k+g_1}\,  e^{ig_1 \cdot \tau_\alpha - ig_2 \cdot (\tau_\beta + \delta)} \cdot \delta_{p-g_1,k-g'_2}
\end{eqnarray}
where we used $t(R) = \frac{1}{N} \sum_q e^{iq\cdot R} t_q \approx \frac{1}{\Omega} \int d^2q \, e^{iq\cdot R} t_q$, and  ${\cal G}$ and ${\cal G}'$ are sets of reciprocal lattice vectors for original bottom and top lattices, respectively. In principle, for a different gauge choice $t(\bR)$ depends on $\alpha,\beta$ indices as well, but since we took the gauge choice where all phase structures for $2p_z$ orbitals are the same, the dependence will be trivial.  In the last line, we just did change of variables. Now we can see that it is nonvanishing only when
\begin{equation}
    p-k = g_1 - {\cal R}_\theta g_2 \quad \textrm{for some reciprocal lattice vectors } g_1, g_2 \in {\cal G}
\end{equation}
Considering that $t_q$ is decreasing fast with $q$, we can only retain most relevant terms where $k+g_1$ is minimized. In terms of a momentum relative to the $\bK^{t,b}_+$ points, we have
\begin{equation}
        (p-\bK_+^t) - (k-\bK_+^b) = (g_1  - {\cal R}_\theta g_2) + q_1.
\end{equation}
Naively, when we consider $k + g_1$, since $k$ does not deviate much from $\bK_+^b$-point, the most relevant $t_{k+g_1}$ would be given when $g_1 = g_2 = 0, G_3, -G_1$ so that $\abs{\bK_+^b + g_1}=\abs{\bK_+^b}$. For cases with $g_1 \neq g_2$, $(i)$ the energy difference between top and bottom electrons are very large, and $(ii)$ $t_{k+g_1}$ would be small, and therefore we ignore such cases.  In fact, each of these cases corresponds to when $\tilde{p}-\tilde{k} = q_1, q_2, q_3$:
\begin{eqnarray}
    g_1 = 0 &\longrightarrow& \tilde{p} - \tilde{k} = q_1 \nonumber \\
    g_1 = G_3 &\longrightarrow& \tilde{p} - \tilde{k} = q_1 + G_3 - {\cal R}_\theta G_3 = q_1 + G_1^M = q_2 \nonumber \\
    g_1 = -G_1 &\longrightarrow & \tilde{p} - \tilde{k} = q_1 - G_1 + {\cal R}_\theta G_1 = q_1 + G_2^M = q_3.
\end{eqnarray}
With this understanding, one can write down three hopping matrices as the following:
\begin{equation}
    T(q_1) = \mqty( 1 & 1 \\1 & 1) \quad T(q_2) = \mqty( z & 1\\ z^* & z ) = e^{-i G_3 \cdot \delta} \mqty( 1 & z^*\\ z & 1) \quad T(q_3) = \mqty( z^* & 1 \\ z & z^* ) = e^{i G_1 \cdot \delta} \mqty( 1 & z \\ z^* & 1)
\end{equation}
where $z = e^{2\pi i/3}$ and since  $G_3 \cdot \tau_0 = G_1 \cdot \tau_0 =  2\pi/3$. Due to the fact that $\delta = -\tau_0$ instead of $\tau_0$, the form is slightly different from the TBG case \cite{MacDonald2011,Po2018}. By proper phase redefinition of Bloch states represented by momentum lattices (gauge degrees of freedom for Bloch states), we can absorb $z$ and $z^*$ factors in front of matrices. Thus, the form of the hopping matrices can be simplified. Moreover, an initial displacement between two layers is not important. In this derivation, it is not difficult to notice that diagonal and off-diagonal entries for Moir\'e hopping matrices can be different. If there is an additional spatial modulation with a Moir\'e scale, given differently between AA(BB) and AB(BA) sites in $t(R+\tau_\alpha - R' - \tau_\beta)$, one would obtain a different values for $t_k$, as explained in Ref.~\cite{Moon2013, Koshino2018}. Finally, to obtain an energy spectrum at Moir\'e momentum $k$, one needs to diagonalize the following Hamiltonian with a certain cutoff:
\begin{equation}
    H = \mqty( H_t(k) & T^\dagger(q_i) & ... \\
    T(q_i) & H_b(k + q_i) & ... \\
    ... & ... & ...)
\end{equation}

Now, we want to point out some subtlety for the generic Moir\'e hopping matrix:
\begin{equation}
    T_n = w_0 + w_1 e^{2\pi n \sigma_3 /3} \sigma_1 e^{-2\pi n \sigma_3 /3}
\end{equation}
where the Pauli operator $\sigma_i$ acts on the sublattice basis. This is the form of the hopping term written in Ref.~\cite{MacDonald2011}, where the $\bK$ and $\bK'$ Dirac Hamiltonian was written as\footnote{$H(\bK+\bk)$ in the MacDonald's paper is a Hamiltonian with respect to the $\bK'$-point in our convention.}
\begin{equation}
    H(\bK+\bk) = \hbar v_F \mqty( 0 & k_x - ik_y\\ k_x + ik_y & 0 ) \qquad
    H(\bK'+\bk) = -\hbar v_F \mqty( 0 & k_x + ik_y\\ k_x - ik_y & 0 ).
\end{equation}
Now, imagine we choose a different basis choice, for example multiplying $(-)$ sign for the $B$-sublattices. This is equivalent to apply  $\sigma_3$ transformation to the operators, and as a result, both $H(\bK+\bk)$ and $H(\bK'+\bk')$ would change its sign. This is the basis chosen in Ref.~\cite{Po2018}. Accordingly, interlayer hopping term $T_0$ would change as well, from $w_0 + w_1 \sigma_1$ to $w_0 - w_1 \sigma_1$ and similarly for others.

\section{Interaction Projection and Intervalley Hund's Coupling}
\label{App:Projection}
Here, we provide the details for the procedure of projecting the Coulomb interaction on the isolated flat band and how to derive the intervalley Hund's coupling. The interaction Hamiltonian can be written as
\beq
\H_{\rm int} = \frac{1}{2} \int d\br_1 d\br_2 \rho(\br) V(\br - \br') \rho(\br').
\label{Hint}
\eeq
Here, $\br$ integration is over the whole space not just the unit cell. For the screed Coulomb interaction, $V(\br)$ is given by
\beq
V(\br) = \frac{e^2}{4\pi \epsilon \epsilon_0} \frac{e^{-\kappa |\br|}}{|\br|}.
\eeq
where $\kappa$ denotes the inverse screening length.
The density operator is given by
\beq
\hat \rho(\br) = \sum_{n,n',\sigma,\sigma',\tau,\tau'} c^\dagger_{n,\sigma,\tau}(\br) c_{n',\sigma',\tau'}(\br),
\eeq
where $\sigma$, $\sigma'$ sum over spin states $\uparrow$, $\downarrow$ and $\tau$, $\tau'$ sum over vallyes $\pm$, and $n$ sums over the relevant set of bands. In the following, we will restrict ourselves to the isolated Moir\'e band and drop the band index $n$. Expansion in the Bloch basis is done by writing
\beq
c_{\sigma,\tau}(\br) = \frac{1}{\sqrt{N}} \sum_{\bk \in \rm BZ} \psi_{\sigma,\tau,\bk}(\br) c_{\sigma,\tau}(\bk),
\eeq
where $N$ is the number of momentum point in the first Brillouin zone which equals to the total number of Moir\'e unit cells in system, and $\psi_\bk(\br)$ are the Bloch states satisfying $\psi_\bk(\br + \bR) = e^{i \bk \cdot \bR} \psi_\bk(\br)$ for a given Moir\'e lattice translation $\bR$. We now split the density into intra- and intervalley components
\begin{gather}
\hat \rho(\br) = \sum_{\sigma,\tau} [\hat{\rho}^+_{\sigma,\tau}(\br) + \hat{\rho}^-_{\sigma,\tau}(\br) ], \\
\hat{\rho}^\pm_{\sigma,\tau}(\br)= \frac{1}{N}\!\!\!  \sum_{\bk,\bk' \in \rm BZ} \psi^\dagger_{\tau,\bk}(\br) \psi_{\pm \tau,\bk'}(\br) c^\dagger_{\sigma,\tau}(\bk) c_{\sigma,\pm \tau}(\bk').
\end{gather}
Here, we used the fact that different spin states are orthogonal. 
If valley symmetry is exact, states belonging to different valleys would also be orthogonal leading to a vanishing intervalley density $\rho^-_{\sigma,\tau}$. However, valley symmetry is broken on the scale of $~|K - K'|^{-1}$ leading to a very small intervalley Hund's coupling term. This term can be usually neglected since it is much smaller than the interaction between intravalley densities. Nevertheless, contributions from this term can lift the degeneracy between different broken symmetry states which are otherwise exactly degenerate, which makes it important to include it in our analysis. We note that the Bloch states are generally vectors with some internal index denoting layer, sublattice, etc which means that the combination $\psi^\dagger \psi$ above denotes an inner product in these internal indices.

The Bloch states can be written in terms of the periodic function $u_\bk(\br)$ which can be expanded in a Fourier series in reciprocal lattice vectors $\bG$ leading to
\beq
\psi_{\bk}(\br) = e^{i \bk \cdot \br} u_{\bk}(\br) = \frac{1}{\sqrt{|\Omega|}} \sum_\bG e^{i (\bG + \bk) \cdot \br} u_{\bk}(\bG),
\eeq
where $\bG$ is the Moir\'e reciprocal lattice vector, and $\Omega$ is the area of the Moir\'e unit cell. Here, $u_{\bk}(\bG)$ are normalized such that $\sum_\bG u^\dagger_{\bk}(\bG) u^{ }_{\bk}(\bG) = 1$. In addition, we can choose the gauge such that the Bloch states satisfy
\beq
\label{Periodic}
u_{\bk + \bG_0}(\bG) = u_\bk(\bG + \bG_0).
\eeq
If the band has a non-vanishing Chern number, it is impossible to choose a smooth and periodic gauge and there would be an additional phase factor in front of the RHS \cite{Kohmoto1985}. In this case, the condition (Eq.~\ref{Periodic}) implies a discontinuity of the phase of $u_\bk$ at the Brillouin zone boundaries.


The interacting Hamiltonian in momentum space is given by
\begin{equation} \label{Vq}
    {\cal H}_\textrm{int} = \frac{1}{2\, \textrm{Vol} } \sum_{\bq} \hat{\rho}(\bq) V(\bq) \hat{\rho}(-\bq)
\end{equation}
where $V(\bq) = \int d\br V(\br) e^{-i \bq \br}$ and  $\textrm{Vol} = N \Omega$. We note that the Fourier transform of $\rho^\pm(\br)$ is not restricted to momenta inside the Moir\'e BZ and it should be expressed in terms of a general momentum $\bq$. 
The density $\hat{\rho}(\bq)$ is generally non-periodic in $\bq$ under reciprocal Moir\'e lattice translations since the Bloch states have a non-trivial spatial structure inside the Moir\'e unit cell. Instead, it decays over some momentum scale comparable to the Moir\'e Brillouin zone size. On the other hand, the Bloch states has no structure inside the unit cell of the original bilayer graphene where a tight-binding description of the orbitals was employed. Hence, the density $\hat{\rho}(\bq)$ is periodic under any reciprocal lattice translation for the original system. As a result, $\hat \rho(\bq)$ consists of several identical narrow peaks centered at reciprocal lattice vectors of the original bilayer graphene $\tilde \bG$ for the intravalley density $\rho^+$ or at $K - K' + \tilde \bG$ for the intervalley density $\rho^-$. This poses a problem since it implies that the summation over $\bq$ in Eq.~\ref{Vq} diverges.

To resolve this issue, we notice that the periodicity of $\hat \rho(\bq)$ in reciprocal space for the original lattice is an artifact of the tight-binding approximation, where an atomic orbital is taken to be point-like. If we instead use the actual shape of the Wannier orbital, the density operator $\hat \rho(\bq)$ will decay for momenta larger than a certain cutoff $\Lambda$ which is given by the inverse size of the Wannier orbitals. Rather than attempting to precisely determine the value of $\Lambda$ from the graphene Wannier orbitals, we will consider $\Lambda$ as a phenomenological parameter of the same order as the size of the original Brillouin zone. This will have the effect of restricting the sum over momenta in Eq.~\ref{Vq} to the vicinity of $\bq=0$ for the intravalley density $\hat \rho^+(\bq)$ and the vicinity of $K - K'$ and $R_{\pm 2\pi/3}(K - K')$ for the intervalley density $\hat \rho^- (\bq)$.

Therefore, we restrict ourselves to the vicinity of 0 for $\rho^+(\bq)$ and $K-K'$ (and its rotation related points) for $\rho^-(\bq)$. In the following, we perform Fourier transform in terms of small deviations around these momenta by defining $\rho_{\sigma,\tau}^\alpha(\bq)$ as(note that $c_{\bk + \bG} = c_\bk$):
\beq
\rho^\alpha_{\sigma,\tau}(\bq) \equiv \int_{N \Omega} d\br e^{-i [\bq - \frac{1 - \alpha}{2} (K_\tau - K_{\alpha \tau})] \cdot \br} \rho^\alpha (\br)= \sum_{\bk} \lambda^\alpha_{\tau,\bq}(\bk) c^\dagger_{\sigma,\tau}(\bk) c_{\sigma,\alpha \tau}(\bk + \bq), \qquad \alpha = \pm
\label{eq:rho}
\eeq
Here, we introduced $K_+ = K$ and $K_- = K'$ and we used that $\psi_{\tau,\bk}(\br) = \sum_\bG e^{i(\bk + K_\tau +\bG)\cdot \br} u_{\tau,\bk}(\bG)$. In addition, we introduced the intra- and intervalley form factors defined by
\begin{align}
\lambda^\pm_{\tau,\bq}(\bk) &= 
\sum_{\bk, \bk' \in \textrm{MBZ}} \sum_{\bG, \bG'} \delta_{K_\tau + \bk + \bG + \bq, K_{\pm \tau} + \bk' + \bG'} \cdot u^\dagger_{\tau,\bk} (\bG) u_{\pm \tau, \bk'}(\bG') \nonumber \\
&=\sum_\bG u^\dagger_{\tau,\bk}(\bG) u_{\pm \tau,\bp(\bk + \bq)}(\bG + \bG(\bk + \bq) ) \equiv \langle u_{\tau,\bk} | u_{\pm \tau, \bk + \bq} \rangle 
\label{lambda}
\end{align}
The function $\bp(\bq)$ and $\bG(\bq)$ are defined to give the projection onto the first BZ and the reciprocal lattice vector corresponding to $\bq$, respectively, such that $\bq = \bG(\bq) + \bp(\bq)$. The last equality is important for the numerical implementation because the summation over $\bq$ can go outside the first BZ whereas the numerical calculation is only carried out in the first BZ.

Time-reversal symmetry dictates that
\begin{align}
    u_{\tau,\bk}(\bG) = u^*_{-\tau,-\bk}(-\bG+\bG_0),
\label{eq:gauge_choice_intervalley}
\end{align}
for some reciprocal lattice vector $\bG_0$. This relation can be exploited for the evaluation of form factors. In fact, a direct numerical evaluation gives $\bG_0$ in our setting. 

The form factors satisfy the identities
\begin{gather}
[\lambda^\pm_{\tau,\bq}(\bk)]^* = \lambda^\pm_{\pm \tau,-\bq}(\bk + \bq), \quad \lambda^\pm_{\tau,\bq}(\bk) = [\lambda^\pm_{-\tau,-\bq}(-\bk)]^*, \nonumber \\ \lambda^\pm_{\tau,\bq}(\bk + \bG) = \lambda^\pm_{\tau,\bq}(\bk).
\label{lambdaIdentities}
\end{gather}
The first identity follows from the definition of the form factor, the second from time-reversal symmetry (Eq.~\ref{eq:gauge_choice_intervalley}) and the third from our periodic gauge choice (Eq.~\ref{Periodic}).

Finally, the resulting interaction can be expanded as a sum of four terms: $\rho^+ \rho^+$ containing intravalley densities, $\rho^- \rho^-$ containing intervalley densities and two cross terms $\rho^+ \rho^-$. The latter ones have to vanish since they necessarily involve densities at large momenta $\bq \pm (K_{+} - K_{-})$ (due to the factor $\lambda_\bq$ which is assumed to decay with $\bq$). The $\rho^- \rho^-$ terms is only non-vanishing when $\tau = -\tau'$. In addition, since $\bq$ is much smaller than $|K_{+} - K_{-}|$, we can ignore the $\bq$ dependence in the interaction term and replace it by the constant $V(|K_{+} - K_{-}|)$. Thus, the resulting Hamiltonian consists of two parts
\beq
\H_{\rm int} = \H_0 + \H_J,
\label{HUJ}
\eeq
$\H_0$ contains the coupling between intravalley densities $\rho^+ \rho^+$  whereas $\H_J$ contains the coupling between intervalley densities $\rho^- \rho^-$. They are given explicitly by
\begin{equation}
\H_0 = \frac{V_0}{2N} \sum_{\sigma, \sigma', \tau, \tau',\bq} \sum_{\bk,\bk' \in BZ} v_\bq \lambda^+_{\tau,\bq}(\bk) [\lambda^+_{\tau',\bq}(\bk')]^*  c^\dagger_{ \sigma,\tau}(\bk) c_{\sigma,\tau}(\bk + \bq) c^\dagger_{ \sigma',\tau'}(\bk' + \bq) c_{\sigma',\tau'}(\bk'),
\label{HU}
\end{equation}
\begin{equation}
\H_J = \frac{3J}{2N} \sum_{\sigma, \sigma', \tau, \bq} \sum_{\bk,\bk' \in BZ}  \lambda^-_{\tau,\bq}(\bk) [\lambda^-_{ -\tau,-\bq}(\bk' + \bq)]^*  c^\dagger_{\sigma,\tau}(\bk) c_{\sigma,-\tau}(\bk + \bq) c^\dagger_{ \sigma',-\tau}(\bk' + \bq) c_{\sigma',\tau}(\bk'),
\label{HJ}
\end{equation}
where the intravalley and intervalley form factors $\lambda^\pm_{\tau,\bq}(\bk)$ are defined as 
\beq
\label{lambda}
\lambda^\pm_{\tau,\bq}(\bk) = \langle u_{\tau,\bk} | u_{\pm \tau, \bk + \bq} \rangle.
\eeq

 All momenta in Eq.~\ref{HU} and Eq.~\ref{HJ} are measured in units of $q_M = \frac{4\pi \theta}{3 \sqrt{3} a}$ with $v_\bq=\abs{q_M}/\sqrt{\bq^2 + \kappa^2}$ denoting the dimensionless screened Coulomb interaction with a screening length $1/\kappa$. The main source of screening is from the gate, which has the distance about 30-50 nm from the sample. The distance is comparable to the Moir\'e length scale, implying that the screening length can be important. In the following calculation, we would use $\kappa = 5\times 10^{7} \textrm{ m}^{-1}$. Rough estimations for $V_0$ and $J$ provide the scale of the two interaction terms and are given by 
\begin{gather}
V_0 = \frac{e^2}{2 \epsilon \epsilon_0 |\Omega| q_M} = \frac{e^2 \theta}{4 \pi \epsilon \epsilon_0 a} \approx 176 \frac{\theta^o}{\epsilon} {\rm meV}, \nonumber \\ J = \frac{e^2}{2 \epsilon \epsilon_0 |\Omega| |K - K'|} = \frac{e^2 \theta^2}{4 \pi \epsilon \epsilon_0 a} \approx 3.1 \frac{(\theta^o)^2}{\epsilon} {\rm meV}.
\end{gather}
Here, we used $|\Omega| = \frac{3 \sqrt{3} a^2}{2\theta^2}$ and used $\theta^o$ to denote the value of $\theta$ in degrees. Using a value of $\epsilon$ of about 5 at twist angles around $1^o$ yields $V_0 \approx 35$ meV and $J = 0.6$ meV. We see that the $J$ term is significantly smaller than the $V_0$ term. It can be important, however, since it identifies the two separate spin-rotation symmetry for $\bK_\pm$ valleys $\textrm{SU(2)}_+\times \textrm{SU(2)}_-$ down to the single spin-rotation $\textrm{SU(2)}$ symmetry, while preserving valley U(1) symmetry. Thus, it can lift the degeneracy between some symmetry breaking states which are degenerate on the level of the $V_0$ interaction. The $J$ term generally has the effect of favoring spin alignment and can be written in the form of inter-valley Hund's coupling as in \cite{Zhang2018}.

We notice that the interaction term is invariant under the gauge transformation
\begin{gather}
c_{\sigma,\tau}(\bk) \rightarrow e^{i \theta_{\tau}(\bk)} c_{\sigma,\tau}(\bk), \\ \lambda_{\tau,\bq}(\bk) \rightarrow e^{i(  \theta_{\tau}(\bk) - \theta_{\tau}(\bk + \bq))} \lambda_{\tau,\bq}(\bk).
\label{Gauge}
\end{gather}
Time-reversal symmetry imposes an additional constraint on the gauge transformation, $\theta_\tau(\bk) = -\theta_\tau(-\bk)$.

\section{Hartree-Fock calculation}
\label{App:HartreeFock}
Here, we provide the details for the Hartree-Fock calculation. Throughout this section, we neglect the Hund's coupling term which is discussed in the previous section and drop the superscript $\pm$ from the form factor $\lambda$ such that $\lambda_{\tau, \bq}(\bk) = \lambda^+_{\tau,\bq}(\bk)$ since we only consider the intravalley form factor here.

We now move on to the general setup for the Hartree-Fock mean field theory. Define the expectation value
\beq
M_{\sigma\tau, \sigma' \tau'}(\bk,\bk') = \langle c^\dagger_{\sigma, \tau}(\bk) c_{ \sigma',\tau'}(\bk') \rangle,
\eeq
which we will assume to be diagonal in $\bk$ and $\bk'$,  $M(\bk, \bk') = \delta_{\bk, \bk'} M(\bk)$. In the following, we will introduce the combined index $\alpha = (\sigma, \tau)$ such that $M(\bk)$ is a matrix with components $M_{\alpha, \alpha'}(\bk)$. Next, we expand the interaction (Eq.~\ref{HUJ}) in the difference $c_\alpha^\dagger c_{\alpha'} - M_{\alpha,\alpha'}$ and neglect terms beyond linear order.

The resulting mean field Hamiltonian has the form
\begin{gather}
    \H_{\rm MF} = \H_K + \H_V, \nonumber \\
    \H_K =  \sum_\bk c_\bk^\dagger [\xi(\bk) + h_0(\bk) + h_1(\bk)] c_\bk , \nonumber \\ \H_V = - \frac{1}{2} \sum_\bk\tr [ h_0(\bk) + h_1(\bk)] M^T(\bk).
    \label{HMF}
\end{gather}
Here, $c_\bk$ is a column vector in the index $\alpha$, $\xi(\bk)$ is a diagonal matrix containing the single particle energies $\xi_{\uparrow/\downarrow, \pm}(\bk)$ and $h_{0,1}(\bk)$ are 4 $\times$ 4 matrices in $\alpha$ given by
\begin{equation}
h_0 =  \frac{V_0}{N} \sum_{\bG, \bk'} \left\{v_\bG \Lambda^+_\bG(\bk) \tr M(\bk') [\Lambda^+_\bG(\bk')]^*  - v_{\bG + \bk'} \Lambda^+_{\bk' + \bG}(\bk) M^T(\bk + \bk') [\Lambda^+_{\bk' + \bG}(\bk)]^*  \right\},
\label{hU}
\end{equation}
and
\begin{equation}
h_1 =  \frac{3J}{N} \sum_{\bG, \bk', \tau} \left\{ P_\tau \Lambda^-_\bG(\bk) \tau_x \tr P_{-\tau} \Lambda^-_{-\bG}(\bk') \tau_x M^T(\bk')]  - P_\tau \Lambda^-_{\bk' + \bG}(\bk) \tau_x M^T(\bk + \bk') P_{-\tau} [\Lambda^-_{-\bk' - \bG}(-\bk)]^T \tau_x \right\}.
\label{hJ}
\end{equation}
 The matrix $\Lambda^\pm_\bq(\bk)$ simply contains the form factors defined in Eq.~\ref{lambda}
\beq
[\Lambda^\pm_\bq(\bk)]_{\alpha,\alpha'} = \delta_{\sigma,\sigma'} \delta_{\tau, \tau'} \lambda^\pm_{\tau, \bq}(\bk),
\eeq
and $P_\pm = \frac{1}{2}(1 \pm \tau_z)$ is the projector on the $\pm$ valley with $\tau_{x,y,z}$ denoting the Pauli matrices in the valley space. 

In both Eq.~\ref{hU} and Eq.~\ref{hJ}, the first term is a Hartree term whereas the second is a Fock term. Hartree terms were neglected in some of the previous mean-field studies \cite{Po2018, Zhang2018} since they are expected to couple only to the density which is determined by the filling in the gapped phase and is independent of the symmetry-breaking order. This is, however, not true in the presence of the form factors which are not the same for the two valleys $\lambda^\pm_{+,\bq}(\bk) \neq \lambda^\pm_{-,\bq}(\bk)$. As a result, the Hartree-term also couples to the valley density and it cannot be neglected. 

It is important here to point out one major difference between our approach and the one employed recently in a self-consistent Hartree-Fock mean field study in twisted bilayer graphen \cite{Xie2018}. In that work, the Hartree-Fock corrections to the flat bands coming from all other ($\sim 150$) bands were taken into account. Here, we will instead make the assumption that the effect of the Hartree-Fock contributions from the other bands is already included at some level in the model parameters which should be either fit to experiments or obtained from {\it ab initio} studies at charge neutrality \cite{Jeil2014, Carr2019}. Thus, we only include the effects arising from filling the isolated band.

To write the self-consistency condition, we diagonalize $h_0(\bk) + h_1(\bk)$ by introducing the variables $d_\bk = U_\bk c_\bk$ for some unitary $U_\bk$. We then impose the  constraint $M_{\alpha,\alpha'}(\bk) = \langle c_\alpha^\dagger(\bk) c_{\alpha'}(\bk) \rangle$. In the following, we will only consider possible gapped phases at integer fillings $\nu$. In this case, the self-consistency condition has the form
\beq
M(\bk) = U^T_\bk \chi U^*_\bk,
\label{MSC}
\eeq
where $\chi$ is a $\bk$-independent matrix containing $\nu$ ones along the diagonal and zeroes everywhere else. \change{This means that $M(\bk)$ is a projection operator satisfying
\beq
M(\bk)^2 = M(\bk) = M(\bk)^\dagger, \qquad \tr M(\bk) = \nu
\eeq
} Our assumption that the phase is gapped has to be checked self-consistently by computing the mean field band structure
\beq
\label{epsilonk}
\epsilon_\bk = \xi_\bk + U^\dagger_\bk h_\bk U_\bk,
\eeq
and ensuring that correlation induced gap for filling $\nu$ defined as
\beq
\label{Delta}
\Delta = \min_\bk \epsilon_{\nu+1, \bk} - \max_\bk \epsilon_{\nu, \bk}
\eeq
is positive. Here, we assumed that the mean field bands $\epsilon_{\alpha,\bk}$ are sorted in order of increasing energy. ($\alpha=1,2,3,4$)

We notice that $M(\bk)$ is, in general, not gauge invariant. Instead it transforms as 
\beq
M_{\sigma,\tau;\sigma',\tau'}(\bk) \rightarrow e^{-i( \theta_\sigma(\bk) - \theta_{\sigma'}(\bk))} M_{\sigma,\tau;\sigma',\tau'}(\bk),
\eeq
under the gauge transformation (Eq.~\ref{Gauge}). In the following mean field analysis, we will choose the gauge such that $\theta_-(\bk) = \theta_+(\bk)$ which guarantees the gauge independence of $M(\bk)$.

{\renewcommand{\arraystretch}{1.3}

\begin{table}[t]
\begin{center}
\begin{tabular}{c||c|c}
$\nu=2$ & Example of $M(\bk)$ &\, Sym. Gen. \\
\hline \hline
SP &\,  $(1+ \sigma_{z} \tau_{0})/2 $ \,& $\sigma_{z}$, $\tau_z$, $\tau_x {\cal K}$  \\
\hline
VP &\,  $(1+ \sigma_{0} \tau_{z})/2 $ \,& $\sigma_{z}$, $\sigma_x$, $\tau_x {\cal K}$  \\
\hline
SVL &\,  $(1+ \sigma_{z} \tau_{z})/2 $ \,&\, $\sigma_{z}$, $\tau_z$, $\sigma_x \tau_x \cal K$  \\
\hline
IVC &\,  $(1+  \sigma_0 \tau_x )/2 $ \,& \, $\sigma_{z}$, $\sigma_x$, $\tau_x {\cal K}$  \\
\hline
\,SIVCL\, &\,  $(1+  \sigma_x\tau_x  )/2 $ \,& $\sigma_x$, $\sigma_z \tau_z$, $\tau_x {\cal K}$  \vspace{0.15in}
\\

$\nu=1,3$ & Example of $M(\bk)$ &\, Sym. Gen. \\
\hline \hline
SVP &\,  $\mqty{( 1+ \sigma_{z} \tau_{0})(1+ \sigma_{0} \tau_{z})/4 } $ \,& $\sigma_{z}$, $\tau_z$ \\
\hline
SPIVC &\,  $\mqty{ (1+ \sigma_{z} \tau_{0}) (1+ \sigma_{0}\tau_x)/4}$\, & $\sigma_z$, $\tau_x {\cal K}$  \\
\hline
SVLIVC \, &\,  $\mqty{ (1+ \sigma_{z} \tau_{z}) (1+ \sigma_{x} \tau_x)/4}$\, &\, $\sigma_{z}\tau_z$, $\sigma_x\tau_x \cal K$  \\
\end{tabular}

\caption{\label{tab:orderparam}
\change{ Examples of order parameter $M(\bk)$ and corresponding independent generators of preserved symmetries for all possible translation-symmetric gapped states at half $\nu=2$ and quarter $\nu=1$ fillings. Note that the $M(\bk)$ can take a more general form. For example, in IVC or SIVCL, $\tau_x$ can be replaced by $c_x \tau_x + c_y \tau_y$ with $c_x^2 + c_y^2 = 1$. Also, for any spin-polarized state, $\sigma_z$ can be replaced by any $\vec{\sigma} = \sin \theta \cos \phi \sigma_x + \sin \theta \sin \phi \sigma_y + \cos \theta \sigma_z$. Here, $\tau_x {\cal K}$ is a spinless time-reversal, where $\cal K$ is an anti-unitary symmetry. \emph{Caveat:} For SVLIVC (which is like SVL+SIVCL) state at $\nu=1,3$, only a certain product structure (in this case spin $S_z$-locked SVL and spin $S_x$-locked SIVCL) would be allowed.    }  }
\end{center}
\end{table}

\subsection{Half-filling $\nu = 2$}
\change{To understand the symmetry breaking at $\nu=2$, we notice that the order parameter can be written as
\beq
M(\bk) = \frac{1}{2}(1 + Q(\bk)), \qquad Q(\bk)^2 = 1, \quad \tr Q(\bk) = 0
\label{M2}
\eeq
$Q(\bk)$ can then be expanded in terms of the generators $\sigma_i \tau_j$ as described in the main text. In the absence of inter-valley Hund's coupling, the problem possesses an SU(2)$\times$SU(2) symmetry corresponding to independent spin rotations in each valley which are generated by $\sigma_{x,y,z} \tau_{0,z}$ in addition to U$_V$(1) valley charge conservation generated by $\tau_z$ and time-reversal symmetry given by $\T = \tau_x K$. Inter-valley Hund's coupling further breaks the SU(2)$\times$SU(2) to SU(2) corresponding to overall rotations. The generators can be grouped into 5 categories according to the symmetries they break as summarized in Table I. We notice that all these terms commute or anticommute with the generators of spin rotation $\sigma_{x,z}$, of U$_V$(1) valley-charge conservation $\tau_z$ and with time-reversal symmetry. In fact, when considering possible symmetry broken states in the limit of flat bands and decoupled valleys, we can always restrict ourselves to matrices $Q(\bk)$ which satisfy this requirement (for some choice of the generators of the symmetries). The reason is that such order are always energetically more favorable. To see this, consider a 'mixed' order given by
\beq
\label{QM}
Q(\bk) = \cos \theta Q_1(\bk) + \sin \theta Q_2(\bk), \qquad Q_{1,2}^2 = 1, \quad \{Q_1, Q_2\} = 0
\eeq
The Fock contribution to the mean-field energy is given by
\beq
E_{\rm HF}[Q] = E_0 + \frac{V_0}{N} \sum_{\bG, \bk'} \left\{ v_\bG \tr \Lambda^+_\bG(\bk) M(\bk) \tr M(\bk') [\Lambda^+_\bG(\bk')]^\dagger -  v_{\bG + \bk'} \tr \Lambda^+_{\bk' + \bG}(\bk) Q(\bk + \bk') [\Lambda^+_{\bk' + \bG}(\bk)]^\dagger Q(\bk) \right\}
\eeq
Substituting the mixed order (\ref{QM}), we find that the mixed term containing both $Q_1$ and $Q_2$ has to vanish since there is some symmetry generator which commutes with $Q_1$ and anticommutes with $Q_2$ (note that the form factors are invariant under all symmetries). This implies that
\beq
E_{\rm HF}[\cos \theta Q_1(\bk) + \sin \theta Q_2(\bk)] = E_0 + \cos^2 \theta E_{\rm HF}[Q_1] + \sin^2 \theta E_{\rm HF}[Q_2]
\eeq
Since the Hartree-Fock solutions has to be extrema of the Hartree-Fock energy functional, we conclude that only pure orders which either commute or anticommute with each symmetry generator are possible self-consistent solution. This justifies restricting ourselves to the list of orders provided in Table I in the main text: SP, VP, SVL, IVC, and SIVCL (such order parameters are in general $\bk$-dependent and may have more complicated forms than the ones written in the second column of the figure, but they have to respect the same symmetries).

If we first neglect the intervalley Hund's coupling, we notice that the mean-field energies of the SP and SVL are equal as well as the IVC and SIVCL since they are related by rotating the spin in one of the valleys. Thus, in the following discussion, we can restrict ourselves to VP, SP, and IVC orders.
}
\subsubsection{Valley polarized (VP) state}
A valley polarized states breaks time-reversal but preserves spin rotation and valley charge. Together with the requirement that the order parameter has the form (\ref{M2}), this yields
\beq
M_{\rm VP}(\bk) = \frac{1}{2}\sigma_0 (1 +  \tau_z).
\eeq
The eigenvalues of $h_\bk$ are given by
\begin{gather}
\epsilon_{\sigma, +,\bk} = \xi_{+,\bk} - \frac{V_0}{N} \sum_{\bG, \bk'} \left\{v_{\bG + \bk'} |\lambda_{+, \bk' + \bG}(\bk)|^2 - 2v_\bG \lambda_{+,\bG}(\bk) \lambda^*_{+,\bG}(\bk') \right\},\\
\epsilon_{\sigma, -,\bk} = \xi_{-,\bk} + \frac{2V_0}{N} \sum_{\bG, \bk'} v_\bG \lambda_{ -,\bG}(\bk) \lambda^*_{+,\bG}(\bk').
\end{gather}
We now need to check the correlated gap defined as
\begin{equation}
    \Delta_{\rm VP} \equiv \min_{\bk,\sigma} \epsilon_{\sigma, -,\bk} - \max_{\bk',\sigma'} \epsilon_{\sigma', +,\bk'} > 0,
\end{equation} 
is positive, so that the fully valley polarized state is a proper gapped state.

The total energy of the valley polarized state is obtained by adding the kinetic energy of the filled bands and the potential energy leading to
\beq
E_{\rm VP} = 2\sum_{\bk}\xi_+(\bk) + \frac{2 V_0}{N} \sum_{\bG} v_\bG \Big|\sum_{\bk} \lambda_{+,\bG}(\bk)\Big|^2 - \frac{V_0}{N} \sum_{\bq,\bk} v_{\bq} |\lambda_{+,\bq}(\bk)|^2.
\eeq

\subsubsection{Spin polarized (SP) state}
Next we assume a spin polarized state along the $z$-direction in both valleys. Such state breaks spin-rotation but preserves time-reversal and valley charge conservation with the order parameter given by
\beq
M_{\rm SP}(\bk) = \frac{1}{2} \tau_0 (\sigma_0 + \sigma_z).
\eeq
The energy eigenvalues are
\begin{gather}
\epsilon_{ \uparrow,\tau,\bk} = \xi_{ \tau,\bk} - \frac{V_0}{N} \sum_{\bG, \bk'} \left\{v_{\bG + \bk'} |\lambda_{\tau, \bk' + \bG}(\bk)|^2 - v_\bG \lambda_{\tau,\bG}(\bk) \sum_{\tau'} \lambda^*_{\tau',\bG}(\bk')  \right\},\\
\epsilon_{\downarrow,\tau,\bk} = \xi_{ \tau,\bk} + \frac{V_0}{N} \sum_{\bG} v_\bG \lambda_{\tau,\bG}(\bk) \sum_{\tau',\bk'} \lambda^*_{\tau',\bG}(\bk').
\end{gather}
We also require the gap $\Delta_{\rm SP} =  \min_{\bk,\tau} \epsilon_{\uparrow, \tau, \bk} - \max_{\bk',\tau'} \epsilon_{\uparrow, \tau', \bk'}$ to be positive so that the spin polarized state is a proper gapped state.

The total energy of the spin polarized state is given by
\beq
    E_{\rm SP} = \sum_{\tau, \bk} \xi_\tau(\bk) + \frac{V_0}{2N} \sum_{\bG} v_\bG \Big|\sum_{\tau,\bk} \lambda_{\tau,\bG}(\bk)\Big|^2 - \frac{V_0}{2N} \sum_{\bq,\tau, \bk} v_{\bq} |\lambda_{\tau,\bq}(\bk)|^2.
\eeq

Comparing to the VP state, we find that the two phases have exactly the same ground state energy. This follows from time-reversal symmetry which implies that $\xi_-(\bk) = \xi_+(-\bk)$ and $\lambda_{-,\bG}(\bk) = \lambda_{+,\bG}(-\bk)$ as well as $|\lambda_{-,\bq}(\bk)| = |\lambda_{+,-\bq}(-\bk)|$. Using the relation  $\sum_{\bk} \lambda_{+,\bG}(\bk) = \sum_{\bk} \lambda_{-,\bG}(\bk)$, one can show that the total energies as well as the gaps are the same for VP and SP states.

\subsubsection{Intervalley coherent (IVC) order}
The intervalley coherent order parameter is given by
\beq
\label{MIVC}
M_{\rm IVC}(\bk) = \sigma_0 
\left( \begin{array}{cc} \cos^2 \frac{\theta_\bk}{2} & \frac{1}{2} \sin \theta_\bk e^{-i \phi_\bk} \\ \frac{1}{2} \sin \theta_\bk e^{i \phi_\bk} & \sin^2 \frac{\theta_\bk}{2} \end{array} \right).
\eeq
We note that it is not possible in general to take the fully polarized limit in the $x-y$ plane and at the same time fulfill the self-consistency conditions. Hence, we include a small $z$ valley polarization parametrized by the angle $\theta_\bk$. We notice that the state (Eq.~\ref{MIVC}) will not break time-reversal symmetry provided that $\theta_{-\bk} = \pi - \theta_\bk$ and $\phi_{-\bk} = - \phi_\bk$ which implies that the average valley polarization $\sum_\bk \cos \theta_\bk$ vanishes.

The mean field Hamiltonian has the form 
\beq
h_\bk = \left(\begin{array}{cc} f_\bk + A_\bk & B_\bk \\ B^*_\bk & f_\bk - A_\bk \end{array} \right),
\eeq
with $f_\bk$, $A_\bk$, $B_\bk$ given by
\begin{gather}
f_\bk = \sum_\tau \left\{ \frac{1}{2}\xi_\tau(\bk) -  \frac{V_0}{4N} \sum_{\bq} v_{\bq} |\lambda_{\tau, \bq}(\bk)|^2 (1 + \cos \theta_{\bk + \bq}) \right. + \left. \frac{V_0}{2N} \sum_{\bG, \bk', \tau'} v_\bG \lambda_{\tau, \bG}(\bk) \lambda_{\tau',\bG}(\bk') (1 + \tau' \cos \theta_{\bk'}) \right\},
\label{fk} \\
A_\bk = \sum_\tau \tau \left\{\frac{1}{2} \xi_\tau(\bk)  - \frac{V_0}{4N} \sum_{\bq} v_{\bq} |\lambda_{\tau, \bq}(\bk)|^2 (1 +\tau \cos \theta_{\bk + \bq}) + \frac{V_0}{2N} \sum_{\bG, \bk', \tau'} v_\bG \lambda_{\tau, \bG}(\bk) \lambda_{\tau',\bG}(\bk') (1 + \tau' \cos \theta_{\bk'}) \right\},
\label{Ak} \\
B_\bk = -\frac{V_0}{2N} \sum_{\bq} v_{\bq} \lambda_{+, \bq}(\bk) \lambda^*_{-, \bq}(\bk) \sin \theta_{\bk + \bk'} e^{- i \phi_{\bk + \bk'}}.
\label{Bk}
\end{gather}
The self-consistency condition reads
\beq
\tan \phi_\bk = -\frac{\Im B_\bk}{\Re B_\bk}, \qquad \tan \theta_\bk = -\frac{|B_\bk|}{A_\bk},
\label{SC}
\eeq
where energy eigenvalues are given by
\beq
\label{ek}
\epsilon_\pm(\bk) = f_\bk \pm \sqrt{A_\bk^2 + |B_\bk|^2},
\eeq
with the gap given by $\Delta_{\rm IVC} =  \min_{\bk} \epsilon_{+, \bk} - \max_{\bk'} \epsilon_{-, \bk'}$ which should be positive for a proper gapped phase. The results of the energy competition between the VP/SP phase and the IVC state obtained by numerically solving the self-consistency equation are given in the main text.

\change{
\subsubsection{Effect of intervalley Hund's coupling}
As we have seen above, the three distinct states with spin polarization, valley polarization or spin-valley locking are degenerate in the absence of intervalley Hund's and their energy is always lower than the energy of the valley off-diagonal orders (IVC and SIVCL). In the following, we want to investigate the effect of intervalley Hund's coupling on these three states. Since this term $J$ is much smaller than the main part of the interaction $V_0$, it suffices to compute it for the three valley-diagonal orders since the valley off-diagonal orders are already energetically unfavorable on the level of $V_0$. Substituting in (\ref{hJ}) we find that
\beq
E_J = \begin{cases} - \frac{3J}{N} \sum_{\tau,\bk,\bq} |\lambda^-_{\tau,\bq}(\bk)|^2 &: \text{SP} \\
0 &: \text{VP} \\
\frac{3J}{N} \sum_{\tau,\bk,\bq} |\lambda^-_{\tau,\bq}(\bk)|^2 &: \text{SVL}
\end{cases}
\eeq
}

\subsection{Quarter-filling $\nu = 1$}
\change{At quarter filling, $\nu = 1$ (similarly for $\nu=3$, with some caveats), we can always write the order parameter as
\beq
M(\bk) = \frac{1}{4}(1 + Q_1(\bk))(1 + Q_2(\bk)), \qquad Q_{1,2}(\bk)^2 = 1, \quad \tr Q_{1,2}(\bk) = 0, \quad [Q_1(\bk), Q_2(\bk)] = 0
\eeq
This leads to three distinct possibilities: (i) $Q_1 = \sigma_{x,y,z}\tau_0$ and $Q_2 = \sigma_0 \tau_z$ which corresponds to a spin and valley polarized state, (ii) $Q_1 = \sigma_{x,y,z}\tau_0$ and $Q_2 = \sigma_0 \tau_{x,y}$ which corresponds to a spin-polarized IVC, and (iii) $Q_1 = \sigma_{z}\tau_z$ and $Q_2 = \sigma_x \tau_x, \sigma_y \tau_y$ which correspond to a spin-valley locked IVC.

The SPIVC and SVLIVC are related by a spin rotation in one of the valleys, thus we can focus only on the competition between SVP and SPIVC. Compared to the VP vs IVC states at half-filling these differ by a factor of 2 in the Fock energy and a factor of 4 in the Hartree energy. Since the former is the main deciding factor in the competition between the phases, the results for the gaps and energy difference between SVP and SPIVC at quarter filling are very similar to those between VP and IVC at half-filling.}


\section{Perturbative solution and competition between VP/SP and IVC}
\label{sec:perturbative}
In this section, we would like to discuss the competition between inter-valley coherent order and valley/spin polarized order in a more general setting that is not too sensitive to the details of the model parameters. To this end, it is useful to derive an approximate solution to the self-consistency equations and compute an analytic expression for the energy difference between the IVC phase and the VP/SP phase. 

In order to make progress analytically, we can write the IVC order parameter as
\beq
\theta_\bk = \frac{\pi}{2} + \gamma_\bk, \qquad \phi_\bk = \beta_\bk.
\label{eq:theta_phi}
\eeq 
where $\gamma_\bk \sim \beta_\bk \sim \delta \ll 1$. This approximation can be justified as follows: the starting symmetry of the isolated band is SU(4) which is broken to SU(2) $\times$ SU(2) due to the asymmetry between the two valley in energies and form factors ($\xi_+(\bk) \neq \xi_-(\bk)$, $\lambda_{+,\bq}(\bk) \neq \lambda_{-,\bq}(\bk)$). 
In the following, we will assume that breaking SU(4) to SU(2) $\times$ SU(2) is not very strong so that the deviation from the situation where the valleys are identical is weak. This condition can be written more explicitly as the requirement that $\frac{|\xi_+(\bk) - \xi_-(\bk)|}{V_0} \sim |\lambda_{+,\bq}(\bk) - \lambda_{-,\bq}(\bk)| \sim \delta \ll 1$. The first part is guaranteed by the small bandwidth whereas the second one can be checked numerically and shown to hold at least for most values of $\bk$ and $\bq$. This is equivalent to expanding in time-reversal symmetry breaking terms within each valley. 

The variables $\gamma_\bk$ and $\beta_\bk$ can be obtained by solving a linearized version of the self-consistency equation as follows. We start by expanding $\theta_\bk$ and $\phi_\bk$ in terms of small deviations $\delta$ from a perfect IVC state in the $\tau_x$ as shown in Eq.~\ref{eq:theta_phi}. Substituting in Eq.~\ref{SC} and expanding to leading order in $\delta$ yields the following set of linear equations given by
\beq
\label{adelta}
\gamma_\bk b_\bk - \sum_{\bk'} F_{\bk,\bk'} \gamma_{\bk'} = a_\bk, \qquad
\beta_\bk b_\bk - \sum_{\bk'} F_{\bk,\bk'} \beta_{\bk'} = -\Im b_\bk,
\eeq
where $a_\bk$ is given by
\beq
a_\bk = \sum_\tau \tau \left\{\frac{\xi_\tau(\bk)}{U}   - \frac{1}{2N} \sum_{\bq} v_{\bq} |\lambda_{\tau, \bq}(\bk)|^2  + \frac{1}{N} \sum_{\bG, \bk', \tau'} v_\bG \lambda_{\tau, \bG}(\bk) \lambda_{\tau',\bG}(\bk')\right\},
\eeq
and $F_{\bk,\bk'}$ and $b_\bk$ are given by
\beq
    F_{\bk,\bk'} =  \frac{1}{N} \sum_{\bG} v_{\bG + \bk' - \bk} |\lambda_{+, \bG + \bk' - \bk}(\bk)|^2, \qquad
    b_\bk = \frac{1}{N} \sum_{\bq} v_{\bq} \lambda_{+, \bq}(\bk) \lambda^*_{-, \bq}(\bk).
    \label{Fb}
\eeq
We notice that $a_\bk$ and $\Im b_\bk$ are of order $\delta$. Substituting in the expression for the energy, the energy difference between the IVC state and the SP/VP state can be written (up to second order in $\delta$) as
\begin{equation}
    \frac{E_{\rm IVC} - E_{\rm SP}}{V_0}=  \frac{1}{4N} \sum_{\bk, \bq} v_{\bq} \Big| \lambda_{+, \bq}(\bk) - \lambda_{-, \bq}(\bk)\Big|^2 \\ + \frac{1}{2} \sum_{\bk, \bk'} \beta_\bk F_{\bk,\bk'} \beta_{\bk'} - \frac{1}{2} \sum_\bk b_\bk (\gamma_\bk^2 + \beta_\bk^2),
    \label{Ediff}
\end{equation}
where $F_{\bk,\bk'}$ and $b_\bk$ are defined in (\ref{Fb}). The first term in Eq.~\ref{Ediff} reproduces the non self-consistent Hartree-Fock energies obtained in Ref.~\cite{Zhang2018} in which case VP/SP is always favored to IVC.

The second and third terms are corrections coming from solving the self-consistency condition. It is instructive to reproduce the results of Ref.~\cite{Po2018} which considers the simplified setting where all form factors are taken equal to 1. In addition, $v_\bq$ was taken equal to a constant which is cutoff at large momenta $\bq \sim \Lambda$ yielding the interaction strength $g = \frac{V_0}{2N} \sum_{|\bq|<\Lambda} = \frac{V_0}{2} \sum_{|\bG|<\Lambda}$. In this case,  $\gamma_\bk = \frac{\xi_+(\bk) - \xi_-(\bk)}{2g}$ and $\beta_\bk = 0$ leading to
\beq
E_{\rm IVC} - E_{\rm SP} = -\frac{1}{4 g} \sum_\bk [\xi_+(\bk) - \xi_-(\bk)]^2,
\eeq 
which implies that the IVC phase is energetically favored to the VP/SP phase in agreement with the conclusion of Ref.~\cite{Po2018} \footnote{The result differs by a factor of 2 due to the incorrect way the large $g$ limit was implemented \cite{Po2018}.}.

Our result (Eq.~\ref{Ediff}) interpolates between these two limits with the first two terms favoring spin or valley polarization and the last term favoring intervalley coherence. The competition between SP/VP and IVC is then settled by the details of the band structure, form factors and interaction. We notice that this expression underestimates the energy of the IVC states when the bands have non-zero Chern number. In this case, it was shown in Ref.~\cite{Bultinck19} that vortices in the IVC order parameter are unavoidable. The existence of vortices is neglected in the expansion (Eq.~\ref{eq:theta_phi}) which assumes that $\phi_\bk$ is small everywhere. This implies that the expression (Eq.~\ref{Ediff}) underestimates the IVC ground state energy for non-zero Chern number.

In order to gain some insights about what parameters control this competition, let us consider a very simplified setting where the Berry curvature is uniform in momentum space with the form factor assuming the simple form \cite{Zhang2018}
\beq
\label{SimpleFormFactor}
\lambda_{\pm, \bq}(\bk) = e^{-\frac{\alpha}{4 } \bq^2 \pm i \frac{B}{2} \bk \wedge \bq}, \qquad B = \frac{2 \pi C}{|A_{\rm BZ}|},
\eeq
Here, $C$ is the Chern number for the $+$ valley and the parameter $\alpha$ determines how quickly the form factor decays with $\bq$ which we take equal to ${2\pi}/{A_{\rm BZ}}$ to reproduce the Landau level form factors for $C=1$. These form factors would be obtained in a Landau level if it is folded into a Brillouin zone of the lattice with the flux density one \cite{usov1988theory}. In addition, we will consider a very simple form of the dispersion corresponding to nearest neighbour tight-binding model on a triangular lattice with hopping amplitude $t e^{\pm i \phi}$ for the $\pm$ valleys. For $C=0$, we know that it is possible to write such a tight-binding model. For non-zero $C$, it is generally impossible to write such tight binding model. However, we can still use the same resulting dispersion and assume that the non-zero Chern number only affects the form factors. This will enable us to disentangle the effects of the band dispersion from those related to band topology.

For the form factors given in Eq.~\ref{SimpleFormFactor}, the self-consistency equations can be solved by performing Fourier transform to real space. Following a series of straightforward steps, we get 
\begin{equation}
\frac{E_{\rm IVC} - E_{\rm SP}}{V_0 N} 
\propto \left\{ \frac{1}{N} \sum_\bk \left[ 1 - e^{-\frac{B^2 \bk^2}{2 \alpha}} I_0\left( \frac{B^2 \bk^2}{2 \alpha} \right) \right]  - \frac{3 \alpha A_{\rm BZ}^2  \eta^2}{2\pi^3\left(1 - e^{-\frac{8 \pi^2}{9 \alpha}} I_0\left( \frac{8\pi^2}{9 \alpha} \right)\right)^2}  \right\} 
\label{Ediffs}
\end{equation}
where $I_b(x)$ is the modified Bessel function of the first kind and $\eta = (t/V_0) \sin \phi$. The proportionality here indicates that we have dropped a constant positive factor given by $\frac{\pi}{|A_{\rm BZ}|} \sqrt{\frac{\pi}{\alpha}}$ which does not influence the competition between the two phases.

The expression (Eq.~\ref{Ediffs}) depends only on two dimensionless parameters: (i) the Chern number $C$ and (ii) $\eta$ which measures the bandwidth relative to the interaction strength multiplied by the strength of time-reversal symmetry breaking within each valley. The first term in Eq.~\ref{Ediffs} is always positive and favors SP/VP state. It vanishes for zero Chern number and increases as the Chern number increases. This suggests that increasing the Chern number favors valley/spin polarization over intervalley coherent order. The second term, on the other hand, favors IVC and increases with increasing the bandwidth or the time-reversal symmetry breaking within each valley. 

\begin{figure}
    \centering
    \includegraphics[width = 0.45 \textwidth]{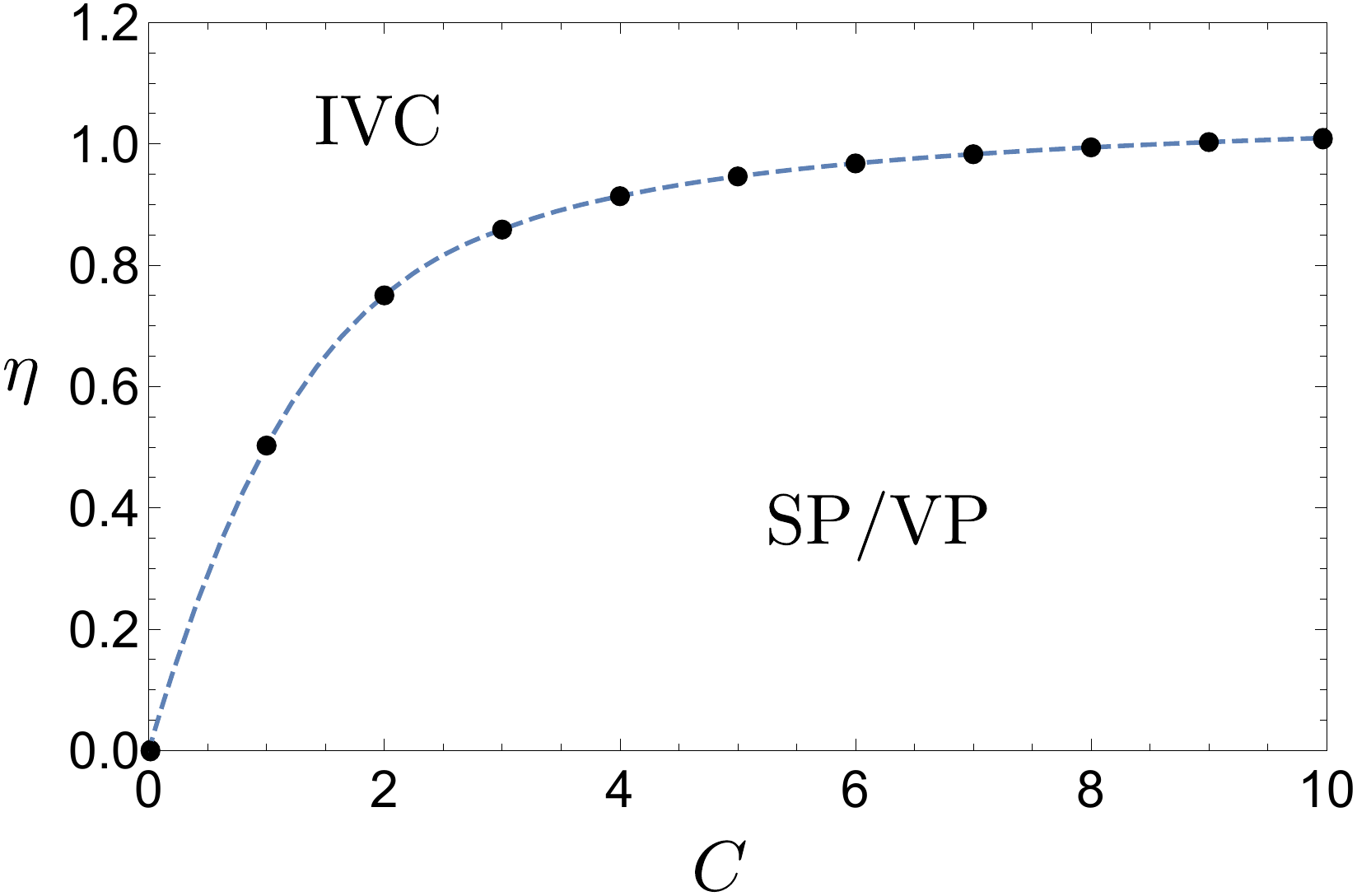}
    \caption{Illustration of the phase diagram obtained from the perturbative solution to the self-consistency equation with the simplified form factor (Eq.~\ref{SimpleFormFactor}) as a function of Chern number and $\eta = (t/V_0) \sin \phi$ which corresponds to the product of the bandwidth (relative to the Coulomb scale) and time-reversal symmetry breaking within each valley. We can see that any non-zero Chern number favors valley or spin polarization over intervalley coherent order as long as the bandwidth is not very large.}
    \label{fig:SPvsIVC}
\end{figure}

The phase diagram for different values of $C$ and $\eta$ is given in Fig.~\ref{fig:SPvsIVC}. For $C=0$, IVC order always wins. This is an artifact of our simple choice for the form factors which corresponds to uniform Berry curvature. In a more realistic model where the Berry curvature vanishes on average but does not vanish everywhere, we expect some region of VP/SP. This is expected to be particularly manifest in the vicinity of topological phase transitions where the valley Chern number changes leading to a large concentration of the Berry curvature at some momenta. For $C \neq 0$, we find that VP/SP is always favored for relatively small values of the bandwidth whereas IVC is favored for relatively large values. Since our approach underestimates the IVC energy for non-zero Chern number (since it ignores vortices \cite{Bultinck19}), we expect the transition from VP/SP to IVC to happen at even larger values of $\eta$ implying that VP/SP is the most energetically favorable insulator at half-filling whenever the bandwidth is relatively narrow.

\section{Spin-triplet superconductivity}
In the following, we provide some details on the discussion related to spin-triplet superconductivity in the main text. The interaction term \beq
\label{HF}
\H = \sum_{\bk, \tau, \sigma} c_{\sigma, \tau, \bk}^\dagger \xi_{\sigma, \tau, \bk} c_{\sigma, \tau, \bk} - g \sum_\bq \bS_\bq \cdot \bS_{-\bq}
\eeq
can be rewritten as
\beq
 g_{\alpha, \beta; \gamma, \delta} \sum_{\bk, \bk', \tau, \tau'} c^\dagger_{\alpha, \tau, \bk + \bq} c^\dagger_{\gamma, \tau', \bk' - \bq} c_{\beta, \tau, \bk} c_{\delta, \tau', \bk'} 
 \eeq
 with $g_{\alpha, \beta; \gamma, \delta} = \sum_a \sigma_{\alpha \beta}^a \sigma_{\gamma \delta}^a$. When performing the BCS decoupling, we restrict ourselves to pairing between time-reversed pairing which corresponds to $\bk' = -\bk$ and $\tau' = -\tau$. In this case, we can define the gap function $\Delta_{\alpha, \beta, \tau,\tau' \bk}=\delta_{\tau, -\tau'} \langle c^\dagger_{\alpha, \tau, \bk} c^\dagger_{\beta, \tau', -\bk} \rangle$, which satisfies the linearized BCS equation 
\beq
\sum_{\bk' \in \rm FS} v_F(\bk')^{-1} g_{\alpha, \beta; \gamma, \delta} \Delta_{\beta, \delta, \tau, \tau', \bk'} = \lambda \Delta_{\alpha, \gamma, \tau, \tau', \bk}
\label{LGE}
\eeq
where $v_F(\bk)$ is the Fermi velocity at point $\bk$ on the Fermi s surface $v_F(\bk) = |\nabla_\bk \epsilon_\bk|$. Choosing $\Delta_\bk$ to be $\bk$-independent, we can simplify (Eq.~\ref{LGE})
\beq
\label{Gap}
\bsigma \cdot (\Delta \bsigma^T) = \tilde \lambda \Delta
\eeq
where $\tilde \lambda$ is related to $\lambda$ by some constant rescaling (coming from the Fermi surface integral), $\bsigma$ is the Pauli matrix vector in spin space and $\Delta$ is a matrix in spin and valley spaces. As discussed in the main text, intervalley pairing is proportional to $\tau_x$ or $\tau_y$ which corresponds to valley triplet or singlet respectively, which, due to the overall antisymmetry of the gap function, implies the former scenario corresponds to a spin-singlet $i \sigma_y$ whereas the latter corresponds to a spin-triplet $i \sigma_y \bd \cdot \bsigma$. Here, $\bd$ is the vector which captures the direction of the spin state. 

The symmetry of the superconducting order parameter is obtained by finding the pairing channel for which $\tilde \lambda$ is positive and maximum. Substituting the spin-singlet and triplet gap functions in (Eq.~\ref{Gap}) yields
\begin{gather}
    \bsigma \cdot (i \sigma_y \tau_x  \bsigma^T) = -3 i \sigma_y \tau_x \rightarrow \tilde \lambda_s = -3 \\
     \bsigma \cdot (i \sigma_y \bd \cdot \bsigma \tau_y  \bsigma^T) =  i \sigma_y \bd \cdot \bsigma \tau_y \rightarrow \tilde \lambda_t = +1 \label{eq:spin_order}
\end{gather}
which implies a valley-singlet spin-triplet superconductor.

\section{Dependence of $T_c$ on magnetic field}

In the following, we will write a simple mean field theory to relate the parameters in the Ginzburg-Landau free energy in Eq.\,6 of the main text to the microscopic parameters. We start by writing the following imaginary time mean-field action
\beq
\label{SMF}
S = \int_0^\beta d\tau \sum_{\bk} \left[ \psi^\dagger_{\bk} (\partial_\tau + \xi_{\bk} + \mu_B \bB \cdot [-\chi \bsigma + \bg_\bk]) \psi_{\bk} + \frac{1}{2} \psi^T_{-\bk} \Delta_\bk \psi_{\bk} + \frac{1}{2} \psi^\dagger_{\bk} \Delta_\bk^\dagger \psi^*_{-\bk}\right] + \frac{\beta}{2g}  \sum_{\bk} \tr \Delta_{\bk} \Delta_{\bk}^\dagger.
\eeq
Here, $\psi$ is a (grassman-valued) spinor in valley and spin spaces, $\sigma$ and $\tau$ are Pauli matrices for the spin and valley degrees of freedom, respectively. $\chi$ is dimensionless magnetic susceptibility and $\Delta$ is a matrix in the valley and spin spaces. Following the discussion of the main text, we take $\Delta_\bk$ to be $\bk$-independent, spin-triplet and valley singlet
\beq
\Delta_\bk = i \sigma_y \bd \cdot \bsigma \tau_y.
\eeq
The magnetic field enters (Eq.~\ref{SMF}) through Zeeman and orbital couplings with the $\bk$-dependent $g$-factor arising form the orbital effect (see the main text). (In (Eq.~\ref{SMF}), $\bg_\bk$ is a diagonal matrix in spin and valley spaces given by $\sigma_0 \diag(\bg_{+,\bk},\bg_{-,\bk})_\tau$). If the parent state is either a weak ferromagnet or close to a ferromagnetic quantum critical point which we anticipate to be the case, then dimensional suscepbtility $\chi$ can be relatively large and cannot be put to 1. 

We can go now to matsubara frequency by writing
\beq
\psi(\tau) = \frac{1}{\sqrt{\beta}} \sum_{\omega_n} e^{i \omega_n \tau} \psi_n, \qquad \omega_n = (2n + 1)\pi/\beta,
\eeq
leading to
\beq
S = \frac{2\beta}{g} \sum_{\bk} \bd_\bk \cdot \bd_\bk^* + \frac{1}{2} \sum_{p=(\omega_n,\bk)} (\psi_p^\dagger \quad \psi^T_{-p}) \left( \begin{matrix} G_p^{-1} - \mu_B \chi \bs \cdot \bB + \mu_B \bg_{\bk} \cdot \bB & \Delta_\bk^\dagger \\ \Delta_\bk & -G_{-p}^{-1} + \mu_B \chi \bs^T \cdot \bB - \mu_B \bg_{-\bk} \cdot \bB \end{matrix} \right) \left( \begin{matrix} \psi_p \\ \psi^*_{-p} \end{matrix} \right).
\eeq
Here, we introduced the Green's function $G_p$ as
\beq
G_p = \frac{1}{i \omega_n + \xi_{\bk}}.
\eeq
where $\xi_{\bk}$ depends on the valley index such that  $\xi_{+,-\bk} = \xi_{-,\bk}$. The fermions can be integrated out leading to a Pfafian which can be written in the exponential as the logarithm of the trace of some operator. The resulting free energy can be expanded in powers of $\bB$ and $\Delta$. 

The term proportional to $\Delta \Delta^\dagger$ provides the standard BCS instability which is given by
\beq
F_{\Delta \Delta^\dagger} = -\frac{1}{2\beta} \sum_{p=(\omega_n,\bk)} \tr \Delta_\bk G_p \Delta_\bk^\dagger G_{-p} = - \frac{2\bd \cdot \bd^*}{\beta} \int d\xi N(\xi) \sum_{\omega_n} \frac{1}{\omega_n^2 + \xi^2} = -\frac{2 \bd \cdot \bd^*}{\beta} \int d\xi N(\xi) f(\xi),
\eeq
where $f(\xi)$ is defined as
\beq
f(\xi) = \sum_{\omega_n} \frac{1}{\omega_n^2 + \xi^2}.
\eeq
The integral over $\xi$ is cut off by the bandwidth $\Lambda$. However, we can choose to perform the $\xi$ integral before the frequency sum in which case, the integral is automatically cut off by $\omega_n$ so that it can be extended to infinity with the cutoff $\Lambda$ moved to the $\omega_n$ sum instead. This leads to
\beq
\int d\xi N(\xi) f(\xi) = \sum_{|\omega_n| < \Lambda} \frac{\pi N(0)}{|\omega_n|} \approx \beta N(0) \int_{1/\beta}^\Lambda d \omega \frac{1}{\omega} = \beta N(0) \log \beta \Lambda.
\eeq
The final result is given by
\beq
F_{\Delta \Delta^\dagger} = -2 N(0) \log \beta \Lambda \, \bd \cdot \bd^*.
\eeq


The term proportional to $\Delta \Delta^\dagger \bB \cdot \bs$ is given by 
\begin{align}
F_{\Delta \Delta^\dagger \bB \cdot \bs} &= -\frac{\mu_B \chi}{2\beta} \sum_{p=(\omega_n,\bk)} \tr \left(\Delta G_p \Delta^\dagger G_{-p} \bB \cdot \bs^T G_{-p} + \Delta G_p \bB \cdot \bs G_{p} \Delta^\dagger G_{-p}\right) \nonumber\\ &=  \frac{4i \mu_B \chi}{\beta} \bB \cdot (\bd^* \times \bd) \int d\xi N(\xi) f'(\xi) = -4i \mu_B \chi N'(0) \log \beta \Lambda \, \bB \cdot (\bd^* \times \bd).
\end{align}
The linear term corresponding to the orbital effect $\Delta \Delta^\dagger \bB \cdot \bg$ vanishes due to time-reversal symmetry which can be seen as follows
\begin{align}
F_{\Delta \Delta^\dagger \bB \cdot \bg} &= -\frac{\mu_B}{2 \beta} \sum_{p=(\omega_n,\bk)} \tr \left(\Delta G_p \Delta^\dagger G_{-p} \tau_z \bB \cdot \bg_{-\bk} G_{-p} + \Delta G_p \tau_z \bB \cdot \bg_{\bk} G_{p} \Delta^\dagger G_{-p}\right) \nonumber\\ 
&=  4\frac{\mu_B \bd \cdot \bd^*}{\beta} \sum_{\tau_{1,2} = \pm} \int_{\rm FS} d\bk (\bB \cdot \bg_{\tau_1, \tau_2 \bk}) \int d\xi N(\xi) f'(\xi) = 0.
\end{align}
The last equality follows from the fact that $\bg_{\tau,\bk}$ is odd under time-reversal symmetry $\bg_{+,-\bk} = - \bg_{-,\bk}$.


The term proportional to $\Delta \Delta^\dagger (\bB \cdot \bs)^2$ is given by
\begin{align}
F_{\Delta \Delta^\dagger (\bB \cdot \bs)^2} &= -\frac{\mu_B^2 \chi^2}{2\beta} \sum_{p=(\omega_n,\bk)} \tr \left(\Delta G_p \Delta^\dagger G_{-p} \bB \cdot \bs^T G_{-p} \bB \cdot \bs^T G_{-p} + \Delta G_p  \bB \cdot \bs G_{p} \Delta^\dagger G_{-p} \bB \cdot \bs^T G_{-p} \right. \nonumber \\ & \left. \qquad \qquad  \qquad \qquad + \Delta G_p  \bB \cdot \bs G_{p} \bB \cdot \bs G_{p} \Delta^\dagger G_{-p} \right) \nonumber \\
&= -2 \frac{\mu_B^2 \chi^2}{\beta} \bB^2 \bd \cdot \bd^* \sum_{p=(\omega_n,\bk)}  (G_p G_{-p}^3 + G_p^2 G_{-p}^2 + G_p^3 G_{-p}) + 4 \frac{\mu_B^2 \chi^2}{\beta} (\bd \cdot \bB) (\bd^* \cdot \bB) \sum_{p=(\omega_n,\bk)}  G_p^2 G_{-p}^2. 
\end{align}
The first term can be simplified by noting that
\beq
\label{T1}
\sum_{p=(\omega_n,\bk)} (G_p G_{-p}^3 + G_p^2 G_{-p}^2 + G_p^3 G_{-p}) = \frac{1}{2} \int d\xi N(\xi) f''(\xi) = \frac{1}{2} N''(0) \beta \log \Lambda \beta,
\eeq
whereas the second term can be evaluated as
\beq
\label{T2}
\sum_{|\omega_n| < \Lambda} \int d\xi \frac{N(\xi)}{(\omega_n^2 + \xi^2)^2} = \frac{\beta}{2} N(0) \int_{1/\beta}^\Lambda d\omega \frac{1}{\omega^3} \approx \frac{\beta^3}{4} N(0),
\eeq
leading to
\beq
F_{\Delta \Delta^\dagger (\bB \cdot \bs)^2} = - \mu_B^2 \chi^2 \bB^2 N''(0) \log \Lambda \beta (\bd \cdot \bd^*) + \mu_B^2 \chi^2 \beta^2 N(0) (\bd \cdot \bB) (\bd^* \cdot \bB).
\eeq

The term proportional to $\Delta \Delta^\dagger (\bB \cdot \bg)^2$ is given by 
\begin{align}
F_{\Delta \Delta^\dagger (\bB \cdot \bg)^2} &= -\frac{\mu_B^2}{2 \beta} \sum_{p=(\omega_n,\bk)} \tr \left(\Delta G_p \Delta^\dagger G_{-p} \bB \cdot \bg_{-\bk} G_{-p} \bB \cdot \bg_{-\bk} G_{-p} + \Delta G_p  \bB \cdot \bg_{\bk} G_{p} \Delta^\dagger G_{-p} \bB \cdot \bg_{-\bk} G_{-p} \right. \nonumber \\ & \left. \qquad \qquad  \qquad \qquad + \Delta G_p  \bB \cdot \bg_{\bk} G_{p} \bB \cdot \bg_{\bk} G_{p} \Delta^\dagger G_{-p} \right) \nonumber \\
& = -2 \frac{\mu_B^2 (\bd \cdot \bd^*)}{\beta} \sum_{p=(\omega_n,\bk)}  \sum_{\tau=\pm} \left[(\bB \cdot \bg_{\tau,-\bk})^2 G_p G_{-p}^3 + (\bB \cdot \bg_{\tau,\bk}) (\bB \cdot \bg_{-\tau,-\bk}) G_p^2 G_{-p}^2 + (\bB \cdot \bg_{\tau,\bk})^2 G_p^3 G_{-p}\right]
\nonumber \\
& =  -2 \frac{\mu_B^2 (\bd \cdot \bd^*)}{\beta} \sum_{p=(\omega_n,\bk)}  \sum_{\tau=\pm} (\bB \cdot \bg_{\tau,\bk})^2  \left[G_p G_{-p}^3 - G_p^2 G_{-p}^2 + G_p^3 G_{-p}\right] \nonumber \\
& = -\mu_B^2 (\bd \cdot \bd^*) (N''(0) \log \Lambda \beta - \beta^2 N(0)) \int_{\rm FS} d\bk \sum_{\tau=\pm} (\bB \cdot \bg_{\tau,\bk})^2.
\end{align}
Here, we used $\bg_{\sigma,\bk} = -\bg_{-\sigma,-\bk}$ to go from the second to the third line and (Eq.~\ref{T1}) and (Eq.~\ref{T2}) to go from the third to the fourth line.

\change{Finally, we evaluate the quartic term $(\Delta^\dagger \Delta)^2$ as
\beq
F_{(\Delta \Delta^\dagger)^2} = \frac{1}{4\beta} \sum_{p=(\omega_n,\bk)} \tr (\Delta G_p \Delta^\dagger G_{-p})^2 = \frac{1}{2\beta} \tr (\bd \cdot \bsigma \, \bd^* \cdot \bsigma)^2 \sum_{p=(\omega_n,\bk)}  G_p^2 G_{-p}^2
\eeq
The summation over $p$ is given by (\ref{T2}), whereas the trace can be evaluated as
\beq
\tr (\bd \cdot \bsigma \, \bd^* \cdot \bsigma)^2 = \tr (\bd \cdot \bd^* + i (\bd \times \bd^*) \cdot \bsigma)^2 = 4 (\bd \cdot \bd^*)^2 - 2 |\bd \cdot \bd|^2
\eeq
leading to
\beq
F_(\Delta \Delta^\dagger)^2 = \frac{\beta^2 N(0)}{4} [2(\bd \cdot \bd^*)^2 - |\bd \cdot \bd|^2]
\eeq}
The Free energy now has the form
\begin{multline}
F = \int_{\rm FS} d\bk \left[ \bd \cdot \bd^* \left(\frac{2}{g} + 2 \mu_B^2 \beta^2 N(0) (\bB \cdot \bg_{+,\bk})^2 - (2N(0) + \mu_B^2 N''(0) (\chi^2 \bB^2 + 2(\bB \cdot \bg_{+,\bk})^2))\log \beta \Lambda \right) \right. 
 \\ \left. + 4i \mu_B \chi \bB \cdot(\bd \times \bd^*) N'(0) \log \beta \Lambda + \mu_B^2 \chi^2 \beta^2 N(0) (\bB \cdot \bd) (\bB \cdot \bd^*) + \frac{\beta^2 N(0)}{4} [2(\bd \cdot \bd^*)^2 - |\bd \cdot \bd|^2] \right] 
\end{multline}
We notice that the second derivative of the density of states can be estimated as $1/\epsilon_F^2$ which is much smaller that $\beta^2$, thus we can throw away all terms containing $N''(0)$. \change{Expanding in $T$ close to $T_c = \Lambda e^{-\frac{1}{g N(0)}}$, we get
\begin{multline}
\label{Fd}
F = \frac{2N(0)}{T_c} \left[ \bd \cdot \bd^* \left(T - T_c + \frac{1}{T_c} \int_{\rm FS} d\bk (\mu_B \bB \cdot \bg_{+,\bk})^2  \right) + 2 i \mu_B \bB \cdot(\bd \times \bd^*) \chi T_c \frac{N'(0)}{N(0)} \right. \log \frac{\Lambda}{T_c}  \\ + \mu_B^2 \chi^2 \frac{1}{2T_c} |\mu_B \bB \cdot \bd|^2 + \frac{1}{8 T_c} [2(\bd \cdot \bd^*)^2 - |\bd \cdot \bd|^2] \bigg]
\end{multline}
Comparing with Eq.~6 in the main text, we find that the coefficients $\kappa, a, b, c, \alpha, \eta$ are given by
\beq
\kappa = \frac{2N(0)}{T_c}, \qquad
a = 2\chi T_c \frac{N'(0)}{N(0)} \ln \frac{\Lambda}{T_c}, \qquad b= \frac{1}{T_c} \int_{\rm FS} d\bk  (\be_{\bB} \cdot \bg_{+,\bk})^2, \qquad c = \frac{\chi^2}{2 T_c}, \qquad \alpha = -2 \eta = \frac{1}{4 T_c}
\eeq}
where $\be_{\bB}$ is the direction of the external magnetic field. We notice that the term $a$ was obtained in the description of the superfluid transition in He${}^3$ \cite{Ambegaokar73}.

}
\end{widetext}

\end{document}